\documentclass[letterpaper,11pt]{article}
\usepackage[margin=1in]{geometry}
\usepackage[margin=0.6in,labelfont=bf,font=small,format=plain,indention=1em]{caption}
\usepackage{titling}
\usepackage{times}
\usepackage{graphicx}
\usepackage{url}
\usepackage[mmddyyyy]{datetime}
\usepackage{enumitem}
\usepackage{multirow}
\usepackage{hhline}

\usepackage[numbers]{natbib}
\makeatletter
\let\@ORIGfnsymbol\@fnsymbol
\renewcommand{\@fnsymbol}[1]{\ensuremath{\ifcase#1\or \dagger}}
\makeatother

\newcommand{\subscript}[1]{\ensuremath{_{\textrm{\scriptsize#1}}}}

\newbox\tempbox
\newdimen\tabledim

\def\tablefont{\fontsize{9}{10}\selectfont}
\def\tabnotefont{\fontsize{9}{10}\selectfont}

\long\def\tbl#1#2{\setbox\tempbox\hbox{\tablefont #2}\tabledim\hsize\advance\tabledim by -\wd\tempbox
\global\divide\tabledim by 2\caption{#1\protect\vphantom{yp}}\vspace{-0.2mm}
\centerline{\box\tempbox}}

\makeatletter
\newenvironment{tabnote}{\par\vskip5pt
\tabnotefont
\@ifnextchar[{\@tabnote}{\@tabnote[]}}{\par}
\def\@tabnote[#1]{\def\@Tempa{#1}\leftskip\tabledim\rightskip\leftskip\ifx\@Tempa\@empty\else{\it #1:}\ \fi\ignorespaces}

\def\tabnoteentry#1#2{\parindent0pt\par{#1}{#2}}
\makeatother

\begin{document}

\setlength{\droptitle}{-4em}
\posttitle{\par\end{center}}
\title{\textbf{\Large Resilient Virtualized Systems Using ReHype}}
\author{Michael Le\thanks{
Currently with IBM T. J. Watson Research Center.}
\ and Yuval Tamir\\
Concurrent Systems Laboratory\\
UCLA Computer Science Department\\
\{mvle,tamir\}@cs.ucla.edu\\
\vspace{-3mm}\\
October 2014}
\date{}
\maketitle

\vspace{-3.5em}
\begin{abstract}
System-level virtualization
introduces critical vulnerabilities
to failures of the software components that implement
virtualization -- the \textit{virtualization infrastructure} (VI).
To mitigate the impact of such failures,
we introduce
a resilient VI (RVI) that can recover individual
VI components from failure, caused by hardware or software faults,
transparently to the hosted virtual machines (VMs).
Much of the focus is on the \textit{ReHype} mechanism
for recovery from hypervisor failures, that
can lead to state corruption
and to inconsistencies
among the states of system components.

ReHype's implementation for the Xen hypervisor
was done incrementally, using fault injection
results to identify
sources of critical corruption and inconsistencies.
This implementation involved 900 LOC,
with memory space overhead of 2.1MB.
Fault injection campaigns, with a variety of fault types,
show that ReHype can successfully recover,
in less than 750ms,
from over 88\% of detected hypervisor failures.

In addition to ReHype,
recovery mechanisms for the other VI components are described.
The overall effectiveness of our RVI
is evaluated hosting a Web service
application, on a cluster of VMs.
With faults in any VI component,
for over 87\% of detected failures,
our recovery mechanisms
allow services provided by the application to
be continuously maintained
despite the resulting failures of VI components.
 \end{abstract}

\section{Introduction}
\label{sec:intro}

System-level virtualization
\cite{Rosenblum05}
enables server consolidation by
allowing multiple virtual machines (VMs) to run
on a single physical host, while providing
workload isolation and flexible resource management.
The virtualization infrastructure (VI)
is comprised of software components
responsible for managing and multiplexing
resources among multiple VMs.
Failure of a VI component due to software bugs
or transient hardware faults generally results
in the failure of the \textit{entire} virtualized system.
Recovery from such a failure typically involves
rebooting the entire system,
resulting in loss of the work in progress
in all the VMs.
This problem can be mitigated through the use of
periodic checkpointing of all the VMs and restoration
of all the VMs to their last checkpoint upon reboot.
However, this involves performance
overhead for checkpointing during normal
operation as well as loss upon recovery
of work done since the last checkpoint.

The hypervisor is the key irreplaceable component of the VI.
The VI typically also includes other components, with
specific functions and labels that vary across different VIs.
The Xen
\cite{Barham03}
VI, that we used as our experimental platform, includes
two other components:
the Privileged VM (PrivVM), that controls, manages,
and coordinates the VMs on the system, and
the Driver VM (DVM)
\cite{Fraser04},
that provides a safe way for I/O devices
to be shared among VMs.
While failure of the PrivVM and/or DVM
can disrupt the ability to manage the system
and/or prevent I/O devices from being accessible to VMs, 
failure of the hypervisor almost immediately
results in the failure of \textit{all} other
system components -- all the VMs and the rest of the VI.

Due to the critical role of the hypervisor,
a large part of this paper focuses on
the design and evaluation of
a mechanism for recovering from hypervisor failures,
using microreboot \cite{Candea04},
called \textit{ReHype}.
Since the operation of VMs can also be disrupted by
the failure of other VI components,
mechanisms for tolerating the failure of
the PrivVM and DVM are also briefly described.
All of these mechanisms together comprise
a resilient VI (RVI) that allows the entire system
to operate through failure of VI components.

ReHype allows VMs to survive hypervisor
failures without any loss of work in progress and
without any performance overhead during normal operation.
To the best of our knowledge, ReHype is the first
mechanism the achieve this.
Upon hypervisor failure, ReHype boots a new
hypervisor instance while preserving the state of running VMs.
VMs are stalled for a short duration during the
hypervisor reboot.
After a new hypervisor is booted, ReHype integrates the preserved
VM states with the new hypervisor to allow the VMs to continue
normal execution.

Failure almost always results in state corruption.
For efficiency, ReHype reuses parts of the vulnerable
state of the failed system,
including the states of all the VMs.
Hence, ReHype,
like any recovery mechanism that relies on vulnerable
state at the time a failure is detected,
cannot be 100\% guaranteed to restore
all system components to valid states.
Furthermore, since ReHype
involves reinitializing part of the hypervisor state while
preserving the rest of the state,
the result of recovery may include
inconsistencies in the hypervisor state,
between hypervisor and VM states,
and between the hypervisor and hardware states.
For example, hypervisor failure can occur in the middle
of handling a hypercall from a VM
or before acknowledging an interrupt from a
device controller.

A key contribution of our work is to
identify the specific sources of state corruptions
and inconsistencies,
determine which of those are most likely to prevent
successful recovery,
and devise mechanisms to overcome these problems.
We have implemented and tested ReHype with the Xen
\cite{Barham03}
hypervisor and VMs running Linux.
We use the results of fault injection to
incrementally enhance
\cite{Ng99}
an initial basic version of ReHype.
These incremental steps improve the rate of successful
recovery from an initial 5.6\% of detected faults
to over 88\% of detected faults.
Our evaluation of the final scheme points to
ways in which the success rate can be further improved.

As discussed further in Section \ref{sec:related},
ReHype builds upon the Otherworld
\cite{Depoutovitch10}
mechanism for microrebooting
the Linux kernel while preserving process states,
and the RootHammer
\cite{Kourai07,Kourai11}
mechanism for rejuvenation of the Xen hypervisor
through a reboot while maintaining VM states in place.
Otherworld microreboots an OS kernel, as opposed to
a hypervisor.
While ReHype does not involve any changes
to the VMs or the applications in the VMs, Otherworld
requires modifications for system call retries and,
in many cases, application-level ``crash procedures'' 
that are invoked upon recovery.
Service interruptions with Otherworld were measured to be
tens of seconds long, as opposed to less than one
second with ReHype.
RootHammer does deal with the Xen hypervisor and provides
proactive rejuvenation.
However, proactive rejuvenation is much simpler
than recovery from failures since it
does not deal with possible arbitrary corruptions and inconsistencies
throughout the system.
Down times with RootHammer were measured to be
tens of seconds long, as opposed to less than one
second with ReHype.

The key contributions of this paper are:
\begin{itemize}[topsep=0mm]
\item
The design and implementation of ReHype --
an efficient and effective mechanism that
enables VMs to survive
across hypervisor failures with minimum interruption.
\item
Using results from fault injection experiments,
identifying specific sources of state corruption
and inconsistencies in the hypervisor, between
hypervisor and VMs, and between the hypervisor and hardware.
The design and implementation of
mechanisms to overcome these problems.
\item
An extensive evaluation of the effectiveness of ReHype
deployed inside a test virtualized environment
and on bare hardware.
The evaluation consists of injecting
hardware and software faults
into the hypervisor while the hypervisor is hosting
para-virtualized and fully-virtualized
VMs running applications.
\item
An analysis and optimization of ReHype's recovery latency.
\item
The design and implementation of a resilient VI (RVI),
that integrates recovery mechanisms for all the VI components.
\item
An evaluation of the resiliency of our complete RVI,
deployed on bare hardware,
hosting a cluster of VMs running the Linux Virtual Server
\cite{Zhang03,Linux}, providing reliable Web service.
\end{itemize}

The following section discusses
the requirements from a resilient hypervisor as well as
key challenges to providing such resiliency
and approaches to meeting these challenges.
Section \ref{sec:basicrec},
describes the implementation
of a version of ReHype that provides
basic transparent hypervisor microreboot but does not
deal with problems caused by state corruptions and inconsistencies.
Incremental improvements to ReHype, based on
fault injection results, are described in Section \ref{sec:improve}.
The details of the experimental setup are presented in
Section \ref{sec:exp}.
Section \ref{sec:ftvi} discusses the
impact on ReHype of incorporating in the hypervisor
support for a PrivVM recovery mechanism.
Additional enhancements to ReHype with respect to
the handling interrupts and VM control are discussed in
Section \ref{sec:enhadd}.
An evaluation of the effectiveness of the final version of ReHype
with respect to different fault types
is presented in Section \ref{sec:eval}.
Sections \ref{sec:validate} and \ref{sec:evalfv}
present, respectively, a validation of
ReHype's effectiveness on bare hardware and
its ability to recover
a hypervisor hosting fully-virtualized (FV) VMs.
The recovery latency of ReHype is
discussed in Section \ref{sec:evallatency}.
Section \ref{sec:rvs} presents the design
and implementation of our complete resilient VI (RVI),
including mechanisms for recovery of the PrivVM and DVM.
An evaluation of the RVI
is presented in Section \ref{sec:evalvi}.
Related work is discussed in Section \ref{sec:related}.
 \section{Tolerating VMM Failure}
\label{sec:vmmfail}

Hardware faults or
software bugs in the virtual machine monitor
(VMM\footnote{
The terms \textit{hypervisor} and \textit{VMM} are used
interchangeably.})
can cause the corruption of VMM state or the state of VMs.
As a result, the VMM or individual VMs
may crash, hang, or perform erroneous actions.
The safest way to recover the system is to reboot
the VMM as well as all of the VMs.
However, this requires a lengthy recovery process
and involves loss of the work in progress of
applications running in the VMs.
Periodic checkpointing of VMs can reduce the
amount of lost work upon recovery.
However, the work
done since the last checkpoint is lost and
there is performance overhead during normal operation
for checkpointing.
The alternative mechanisms discussed below involve less overhead
and lost work but may result in recovery of only
parts of the system or even a complete failure
to recover a working system.
This section discusses the basic design alternatives
for mechanisms that can recover from VMM failure.

Virtualization is often used to consolidate the workloads
of multiple physical servers on a single physical host.
With multiple physical servers, a single software or
transient hardware fault may cause the failure of one of the servers.
An aggressive reliability goal for a virtualized system
is to do no worse than a cluster of physical servers.
Hence, if a transient hardware fault or a software fault in
any component (including the VMM) affects
only one VM running applications, the goal is met.
Recovery from VMM failure that avoids losing work
in progress in the VMs necessarily relies on
utilizing the VM states at the time of failure detection.
One or more of those VM states may be corrupted,
resulting in the failure of those VMs even
if the rest of the system is restored to correct operation.
Based on the reliability goal above, we define
recovery from VMM failure to be successful if no
more than one of the existing VMs running applications
fails \textit{and} the recovered VMM maintains its ability to
host the other existing VMs as well as create
and host new VMs \cite{Le11}.

Successfully ``tolerating'' VMM failure requires
detection of such failures and successfully
recovering from them, as defined above.
To accomplish this goal,
mechanisms must exist to:
(1) detect VMM failure,
(2) repair VMM corruption,
and
(3) resolve inconsistencies within the VMM,
between the VMM and VMs, and between the VMM and the hardware.
As described in Subsection \ref{sec:det},
detecting VMM failure boils down to being
able to detect a VMM crash, hang, or silent corruption.
Subsection \ref{sec:recappr} discusses
different approaches to repairing
VMM corruption and the tradeoffs among them in terms of
implementation complexity and expected success rates.
Inconsistencies
among the states of different components following
recovery may be resolved entirely in the VMM or may require
VM modifications.
Details of the sources of inconsistencies and techniques
for resolving inconsistencies are
described in Subsection \ref{sec:incons}.

\subsection{Detection}
\label{sec:det}

Faults in the VMM can manifest as VMM crashes, hangs, or silent
corruption (arbitrary incorrect actions).
Crashes can be detected using existing VMM panic and exception
handlers --
if the VMM panics, a crash has occurred.
Detecting VMM hangs requires external hardware.
A typical hang detector, such as the one implemented
in the Xen VMM, uses a watchdog timer
that sends periodic interrupts to the VMM.
The interrupt handler checks whether the VMM
has performed certain actions since the last
time the handler was invoked.
If it has not, the handler signals a hang.

Silent VMM corruption is more difficult to detect.
Detection mechanisms involve redundant (e.g., replicated)
data structures and redundant computations
(e.g., performing sanity checks).
Fortunately, our fault injection results (Section \ref{sec:eval})
indicate that the majority of VMM failures (65\%-80\%) 
are caused by crashes and hangs and are thus detectable using
the simple mechanisms discussed above.

\subsection{Repairing VMM Corruption}
\label{sec:recappr}

Repair is initiated
when the detection mechanism invokes a failure handler.
Corrupted VMM state can then be repaired by either
identifying and fixing the specific parts of
the VMM state that are corrupted or simply
booting a new VMM instance.
A major difficulty with the first alternative
is the requirement to identify which parts
of the state are erroneous.
This is likely to require significant overhead for
maintaining redundant data structures.
Furthermore, complex repair operations
performed in the context of a failed VMM
can increase the chance of failed recoveries
\cite{Sullivan91}.
The approach of using nested virtualization
with on-demand checkpointing proposed by
\cite{Tan12}
can potentially be used to
repair VMM state corruption
but incurs a high runtime overhead
(see Section \ref{sec:related}).
Hence, we focus on repair by booting a new VMM instance.

Normally, a full system reboot causes the loss
of all the VM states.
As mentioned earlier, safe recovery from such a reboot
involves high overhead.
To eliminate the overhead during normal operation,
the VM states can be checkpointed to stable storage
only after VMM failure is detected (in the failure handler).
Once a new VMM instance boots up, the VMs can be restored.
However, checkpointing VM states in the context
of a failed VMM increases the chance of failed recoveries,
since the VMM must perform I/O and access possibly corrupted structures
that hold VM states.
In addition, the time to save and restore VM states
results in slow recovery, leading to lengthy service interruptions.

An alternative approach to a system-wide reboot,
is to microreboot
\cite{Candea04}
the VMM.
With this approach, VM states are preserved in memory across the reboot.
This avoids the overhead of checkpointing VM states
to stable storage.
Once the new VMM has been booted, it must
be re-integrated with the preserved VMs.
This re-integration can be done by either recreating
the VMM structures used to manage the VMs
or reusing VMM structures preserved
from the old VMM.
Either way, some amount of VMM data needs to be preserved
across a VMM reboot for the re-integration process.

Variations of the VMM microreboot approach can be
categorized based on two dimensions:
(I) whether the new VMM is rebooted in place (as with ReHype)
or in a reserved memory area
(similarly to Otherworld \cite{Depoutovitch10}); and
(II) whether the VMM structures for managing VMs
from the old VMM instance are preserved and directly reused, or
new instances of these structures are created and
populated with state from the old VMM instance.
The choice in Dimension (I) affects the complexity of
the operations that must be done in the failure handler.
The choice in Dimension (II) affects the complexity of
the operations required for reintegrating
the preserved VMs with the new VMM instance.
Since, in general, minimizing the complexity of operations
required for recovery
increases the probability of successful recovery,
these choices are important.
The rest of this subsection discusses these variations.

If the new VMM is booted in place, the failure
handler must perform two operations
that are not needed if the VMM is booted into
a reserved memory area:
1) preserve VMM state (data structures) from the failed
VMM instance, needed for reintegration with the preserved VMs,
by copying it to a reserved memory area; and
2) overwrite the existing VMM image in memory with a new image.
If the VMM is booted into a reserved memory area,
the entire old VMM state is preserved
since, on boot, the new VMM
is confined to the reserved memory area.
Thus, the copying of old VMM state is not needed.
In addition,
if the VMM is booted into a reserved memory area,
the new VMM image can be preloaded into the reserved
memory area without affecting the operation of the current VMM.
Obviously, this choice involves memory
overhead for the required reserved area.

Since the state of the old VMM instance may be corrupted,
the ability to successfully recover is directly related
to the amount of data reused from the old VMM instance.
In some cases, data structures in the new VMM instance
can be re-initialized to static values (e.g., clearing all locks)
or reconstructed from sources that are unlikely
to be corrupted (e.g., obtaining the CPUID of a core
from the hardware).
However, some data structures are dynamically
updated, based on the activity of the system,
and cannot be re-initialized with static or ``safe'' values.
For example,
a VM's page table or the VMM's timer heap.

With respect to the choices along Dimension (II) above,
reusing preserved data structures
from the old VMM instance is simpler to implement,
as only pointers to the preserved structures need
to be restored in the new VMM instance.
Creating new instances of VMM data structures
is more complex, as it requires
deep copy of all the required structures from the old VMM
and updating all pointers within those structures.
With either of the alternatives along Dimension (II),
there is a possibility of
ending up with corrupted values in the new VMM's
data structures.
With new data structure instances, there is a higher probability
of failure during the deep copy operations in the
reintegration phase.
If the preserved structures are 
reused, there is a greater risk of also introducing
into the new VMM instance corrupted pointers,
which may lead to further corruption later on,
after the system resumes normal operation.
This risk with the reuse of preserved structures can
be partially mitigated by proactively rejuvenating
the system \cite{Kourai11} soon after recovery.

Given the tradeoffs presented in this subsection,
ReHype uses the microreboot approach and
opts for a simple implementation that
does not require major modifications to the VMM.
With ReHype, the VMM is rebooted in place.
ReHype preserves and reuses almost all of the VMM's dynamic
memory, but updates a few key data structures with ``safe'' values,
as described in Sections \ref{sec:basicrec} and \ref{sec:improve}.
The benefit of reusing most of the VMM's data structures
is that it allowed ReHype to be easily integrated
into a VMM (Xen in our case) with minor (900 LOC added/modified)
modifications.

\subsection{Resolving Inconsistencies}
\label{sec:incons}
\hyphenation{VCPU}
\hyphenation{VCPUs}

VMs and VMMs are generally designed and implemented with
the assumption that lower system layers are,
for the most part, reliable.
Hardware mechanisms typically assume that the layer
above will interact with the hardware correctly
(according to the specifications).
These assumptions are violated when the VMM fails due
to hardware or VMM software faults.
Thus, after recovery from a failed VMM, even if none
of the states of the
system components are corrupted, these states may
be \textit{inconsistent}, preventing the system from
operating correctly.

The VMM executes some operations
in critical sections to ensure atomicity,
e.g. updating a VM grant table.
Atomicity can be violated when a VMM failure occurs in the middle
of such critical sections.
In such cases, some data
structures may be partially updated, leading to
inconsistencies within the VMM (VMM/VMM).
The VMs expect the VMM
to provide a monotonically increasing system time,
handle hypercalls, and deliver interrupts.
The hardware expects the VMM to
acknowledge all interrupts it receives.
When a VMM failure occurs, the assumption of a reliable VMM is violated
and this can lead to inconsistencies between the VMM and VMs (VMM/VM)
and between the VMM and hardware (VMM/hardware).
The recovery process must resolve these inconsistencies
so that the virtualized system can continue to operate correctly.
The rest of this subsection
discusses these inconsistencies
and techniques for their resolution.

Sources of VMM/VMM inconsistencies
include partially updated structures,
unreleased locks, and memory leaks.
The options for resolving these inconsistencies are, essentially,
special cases of the options for dealing with state
corruption, discussed in the previous subsection.
Resolving inconsistencies caused by partially updated structures
requires either constructing new instances of
the data structures using information from the failed
VMM, or using redundant information, logged prior to failure,
to fix the preserved instance of the data structure.
A scheduler's run queue is an example of a data structure
for which the former technique can be used.
Inconsistency can occur if a
VCPU becomes runnable
but the VMM fails before inserting it into the run queue.
Resolving this inconsistency requires re-initializing
the run queue to empty upon bootup and
re-inserting all runnable VCPUs (obtained from the failed VMM)
into the run queue.
For other data structures, such as the ones that track memory page
usage information,
reconstruction is more difficult,
so the latter technique may be preferable.
For instance, an inconsistency can occur if a failure
happens right when a page use counter has been updated but
before that page has been added to a page table.
Resolving this inconsistency by traversing all page table entries
to count the actual mappings to that page can be done, but
is complex and slow.
Instead, the entire mapping operation can be made atomic
with respect to failure using \textit{write-ahead logging},
involving a small overhead
during normal operation and simple, fast correction
of any inconsistencies upon recovery.

Locks and semaphores acquired
prior to VMM failure
must be released (re-initialized to a static value)
upon recovery to allow the system to reacquire
them when needed.
In order to do so,
all locks and semaphores must be tracked and
re-initialized in data structures that are reused or
copied from the failed VMM.

A memory leak can occur if a failure happens
between the allocation and freeing of a memory region
in the VMM.
Since failures are rare,
such a memory leak is generally benign, as long as the
leaked region is small relative to the total memory size.
After VMM recovery, the system can be scheduled to be rebooted
to reclaim leaked memory.
Alternatively, leveraging
\cite{Kourai11},
after recovery, the virtualized system can be quickly rejuvenated
to reclaim any leaked memory.

Sources of VMM/VM inconsistency
include erratic
and/or non-monotonic changes in system time,
partially executed hypercalls, and undelivered virtual interrupts.
The correct operation of many VMs depends on a
system time that is monotonically increasing at a constant rate.
In a virtualized system, the VMs' source of time is the VMM.
When a VMM is rebooted, its time keeping structures
are reset, potentially resulting in a time source for VMs
that is erratic (e.g., ceases to advance or
advances suddenly by a significant amount)
or that is not monotonically increasing.
In addition, such a reset can result in
timer events set using
time relative to the VMM's system time prior to recovery
to be delayed.
One technique for resolving this inconsistency
is to simply save the VMM time structures upon failure and restore
those structures after the VMM reboot and before
the VMs are scheduled to run.
This allows time to continue moving forward with no interruption
visible by the VMs.
For external entities that interact with the VMs
and expect time to remain approximately synchronized
with real time, additional mechanisms,
such as NTP, are required to slowly accelerate
time in the virtualized system to catch up with real time.

When the VMM recovers from a failure, partially executed
hypercalls must be re-executed.
Our experimental results show that, at least
for the hypercalls that were exercised by our target system,
simply retrying hypercalls works most of the time
and allows VMs to continue to operate.
However, hypercalls that are not idempotent may fail on
a retry, in which case
the VM executing the hypercall may also fail.

Hypercall retry can be implemented by modifying the VM
to add a ``wrapper'' around hypercall invocation
that will re-invoke the hypercall
if a retry value is returned by the VMM.
The VMM must also be
modified to return, upon recovery, a retry value indicating a
partially executed hypercall.
This approach provides the VMs control
over which hypercalls to retry and allows the VMs
to gracefully fail if a hypercall retry is unsuccessful.

Hypercall retry can also be implemented without modifying the VMs.
To force re-execution of a hypercall after recovery,
the VMM adjusts the VM's instruction
pointer to point back to the hypercall instruction (usually a trapping
instruction).
When the VM is scheduled to run,
the very next instruction it executes
will be the hypercall.
This mechanism is already used in the Xen
\cite{Barham03}
VMM to allow the preemption of
long running hypercalls
transparently to the VMs.
It should be noted that
this mechanism can also deal with VMM failures
that occur while in the middle of handling
a VM trapping instruction that is not part of a hypercall.

The VMM is responsible for delivering interrupts from hardware and
event signals from other VMs as virtual interrupts
to the destination VM.
These virtual interrupts may be lost if the VMM fails.
Some inconsistencies of this type can be resolved
without any modifications to the system
by relying on existing timeout mechanisms that
are implemented in the kernels and
device drivers of the VMs.
A timeout handler can resend commands
to a device or resignal another VM if an expected interrupt
does not arrive within a specified period of time.
We have verified that
timeout mechanisms exist for
the Linux SCSI block driver (used for SATA disks) and the
Intel E1000 NIC driver, representing
the most important devices for servers
(storage and network controllers).
Obviously, such timeout mechanisms
do not deal with lost interrupts from unsolicited sources, such
as packet reception from a network device.
However, at least for network devices,
the problem is ultimately resolved by existing higher-level
end-to-end protocols (e.g., TCP).

A source of VMM/hardware inconsistency
is unacknowledged interrupts.
The processor blocks delivery of
pending interrupts that are of lower or equal priority
than the current interrupt being serviced by the VMM.
These blocked interrupts can be delivered
once the in-service interrupt has been acknowledged by the VMM.
In addition,
for level-triggered interrupts,
the interrupt controllers will block an interrupt source until
the previous interrupt from that source
has been acknowledged.
Since VMM failure can occur at any time,
the interrupt being serviced
at the point of failure
may never
get acknowledged, thus blocking
interrupts of lower or equal priority
indefinitely.
If VMM recovery is done without performing a hardware reset,
a mechanism is needed to either
reset both the pending interrupt state in the processor and
the I/O controller, or
acknowledge all pending interrupts
during recovery.
In the case of acknowledging pending interrupts,
the interrupt source must be blocked at the interrupt
controller before the interrupt is acknowledged
to prevent another interrupt from slipping by before the VMM is ready
to handle the interrupt.
 \section{Transparent VMM Microreboot}
\label{sec:basicrec}

We have implemented a ReHype prototype
for version 3.3.0 of the Xen
\cite{Barham03}
VMM.
This section describes the
implementation of a version of ReHype that provides
the basic capability to microreboot the VMM
while preserving the running VMs and allowing
them to resume normal execution following the microreboot.
Improvements to the basic scheme that enhance
recovery success rates
are discussed in Section \ref{sec:improve}.

To microreboot the VMM, ReHype uses
the existing
Xen port of the Kdump
\cite{Goyal05}
tool.
Kdump is a kernel debugging tool that
provides facilities to allow a crashed system to load
a pristine kernel image, in this case the VMM image,
on top of the existing
image, and directly transfer control to it.
The Kdump tool by itself, however,
does not provide any facilities
to preserve parts of memory, such as those holding
VM states.
The burden of memory preservation is on the kernel or VMM being booted.

A VMM microreboot is differentiated from a normal VMM boot
by the value of a global flag added to the initialized
data segment of Xen.
The flag is clear in the original VMM image on disk.
Upon the initial system boot, the recovery image
is loaded to memory using the Kdump tool and
the flag in that image is then set.
All the modifications to the bootup process described henceforth
refer to
actions performed when the flag is set (the microreboot path).

On boot, the stock Xen VMM 
initializes the entire system memory
and allocates memory for its own use.
The modifications for ReHype must ensure that, upon recovery,
the new VMM instance preserves the memory used by VMs
and memory that was used by the previous VMM instance
to hold state needed for managing the VMs.
Hence, as described in Subsection \ref{sec:micro},
the ReHype version of the Xen VMM
allocates ``around'' the preserved memory regions
during a VMM microreboot.
When the new VMM instance is booted and initialized,
it does not contain
information about the running VMs, and thus
has no way to run and manage them.
Subsection \ref{sec:reintegrate} describes how
the VMM and VM states preserved during VMM recovery
are re-integrated with the new VMM instance.

\subsection{Preserving VMM and VM States}
\label{sec:micro}

The state that must be preserved across a VMM microreboot includes
information in the VMM's static data segments,
the VMM's heap,
the structure in the VMM that holds information about each machine page,
and special segments that are normally used only
during VMM bootup.

VMM microreboot involves
overwriting the existing VMM image
(code, initialized data, and bss) with
a pristine image.
The VMM's static data segments (initialized data and bss)
contain critical information needed for the
operation of the system following bootup of the new VMM instance.
For example, this includes interrupt descriptors
and pointers to structures on the Xen heap, such
as Domain 0's VM
structure and the list of domains.
Hence, some of the information in
the old static data segments must be preserved.
While it is only necessary to preserve a subset of the static
data segments, since they are relatively small,
we reduce the implementation complexity by
preserving the segments in their entirety.
\begin{figure}
\centerline{\includegraphics[width=4in]{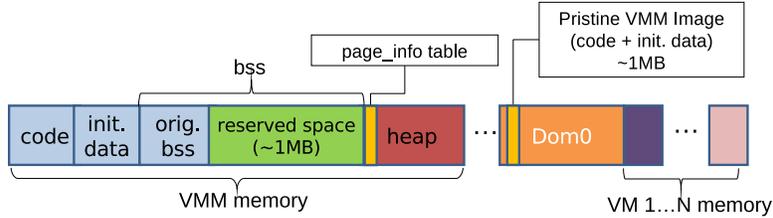}}
\caption{Layout of system memory for ReHype.}
\label{fig:memlayout}
\end{figure}

Figure \ref{fig:memlayout} shows the memory layout for
Xen, as modified for ReHype.
In particular, the bss segment is extended by approximately 1MB ---
sufficient space to hold complete copies of the
original bss and initialized data segments.
This is done by changing the linker script used for creating
the Xen image.
Since the area reserved is in the bss segment, no extra disk space
is taken up for the new VMM image.
The bss segment only takes up space when it is loaded into memory.
Upon failure detection, before initiating a VMM microreboot,
the failure handler copies
the initialized data and (unextended) bss segments
to this new reserved memory.

As discussed above,
for each VM, the state that must be preserved includes
both the VM's memory image and parts of the VMM state
used to manage the VM.
Since this information is maintained on the VMM's heap,
the VMM's heap must be preserved.
Preserving the VMM's heap requires
modifications of the VMM's bootup code.
During VMM initialization, before the heap is created,
the old heap (from the previous VMM instance)
is walked to identify all the free pages.
When the new heap is created and populated with free pages,
only pages that are free in the old heap
are added to the new heap.
This ensures that the new VMM will not allocate (overwrite)
any pages that are still in use.
To access the old heap,
the page table from the old VMM must be restored.
This requires copying the old page directory
from the old bss segment, preserved as discussed above,
to the new bss segment.

The Xen VMM maintains a structure (page\_info)
that holds information about each machine page,
such as the page ownership and use count.
For all the pages that are preserved across a VMM microreboot,
the information in this structure must be preserved.
This structure is allocated in a special memory area,
between the bss segment and the heap.
The VMM bootup code was modified to avoid initializing
the page\_info entries for pages that are not free.

The stock Xen VMM image includes two static segments (init.text
and init.data) that are normally used during bootup
and then freed to the heap
so that they can be used for other purposes.
Hence, with stock Xen, a microreboot would overwrite
these segments, potentially corrupting data in pages
that had been reallocated.
To prevent this problem, the bootup code
(normal and microreboot)
has been modified to
avoid freeing these pages.
This results is an extra 100KB memory overhead.

Preserving the heap and static data segments of a failed VMM
is unsafe --- it can result in recovery failure if those
preserved values are corrupted by the failed VMM.
Section \ref{sec:improve} discusses mechanisms that
dramatically improve the chances of successful
recoveries despite re-using the preserved heap and static data segments.

\subsection{Re-integrating VMM and VM States}
\label{sec:reintegrate}

Following a VMM microreboot,
the VMM does not have the information
required to resume execution and manage the VMs that
were running at the time of failure.
The missing system state
includes the list of VMs it was managing,
the system time that was provided to the VMs,
information for interrupt routing (to processors) and forwarding (to VMs),
and timer events that were scheduled by the VMs.
To allow the virtualized system to continue running,
these components of the system state must be restored.
As discussed in the previous subsection,
all the required state is preserved across a VMM microreboot.
Hence, all that is needed is to re-integrate the
preserved information with the new VMM instance.
This re-integration is accomplished by
copying a few key values from the old static data segments
to the new static data segments.
The restoration is done before the VMs can be scheduled to run.
The following structures are restored:
\begin{itemize}
\item
Pointer to xmalloc free list:
prevent memory leaks.
\item
Pointers to the domain list and hash table:
allow Xen to access the state of the running VMs.
\item
Pointer to the Domain 0 descriptor:
since Domain 0
is not rebooted as part of recovery,
the pointer to it must be restored to allow Xen
access to the Domain\ 0 structure.
\item
Pointers to timer event objects:
restore pending timer events on the old timer heap
to the new timer heap.
\item
Pointer to the machine-to-physical (m2p) mapping table:
make available mapping of machine frame numbers
to physical frame numbers.
\item
System time variables:
maintain monotonically increasing time.
The time-stamp counter (TSC) must not be reset.
\item
IRQ descriptor table and IO-APIC entries, including
correct IRQ routing affinity and mask:
allow VMs to continue to receive interrupts from their devices.
\item
Structures for tracking the mappings of
pages shared between VMs and the VMM:
prevents overwriting mappings that are still in use.
\end{itemize}

There are two additional differences between the
VMM microreboot path and a normal VMM boot:
\mbox{Domain 0} is not created and VMs are re-associated with
the scheduler.
The bootup code has been modified to skip Domain 0 creation
and to restore the global pointer to Domain 0
so that the new Xen can access Domain 0's state.
VMs are re-associated with the VMM's scheduler by
invoking the scheduling initialization routines
for each VM and
inserting runnable VCPUs into the new run queue.

Some of the structures needed by the VMM
are re-created during a microreboot.
These include the idle domain as well as structures holding
hardware information, such as the model and type of the
CPU and the amount of memory available.
For structures that are re-created on the heap,
ReHype prevents a memory leak by first freeing the
old structures.
 \section{VMM Recovery Improvements}
\label{sec:improve}

The scheme presented in the previous section
provides basic capabilities for VMM microreboot.
However, as explained below, with this basic mechanism
the probability of successful recovery is very low.
This section starts with the basic scheme and incrementally
enhances it to achieve high recovery success rates.
Table \ref{table:rec} shows the mechanisms used to improve
the basic recovery scheme.
As in 
\cite{Ng99},
the choice of enhancements is guided
by results from fault injection experiments.
The last two mechanisms in Table \ref{table:rec}
were motivated by additional experimentation
and are discussed in more detail
in Section \ref{sec:enhadd}.

\begin{table}
\tbl{
Improvements over the basic ReHype recovery.
\label{table:rec}}{
\begin{tabular}{|p{3.7cm}|p{6cm}|}
\hline
\multicolumn{1}{|c|}{Mechanism} & \multicolumn{1}{c|}{Description} \\
\hhline{|=|=|}
NMI IPI &
Use NMI IPIs in failure handler.
Avoid IPI blocking by failed VMM. \\
\hline
Acknowledge interrupts &
Acknowledges all
in-service
interrupts in all processors to
avoid blocked interrupts after recovery. \\
\hline
Hypercall retry &
All partially executed hypercalls are retried transparently to the VMs. \\
\hline
FixSP &
Stack pointer set to ``safe'' value in failure handler. \\
\hline
NMI ``ack'' &
Execute iret to ``ack'' NMI when hang detected on non-CPU0. \\
\hline
Reinitialize locks &
Dynamically allocated spin locks
and non-spin locks are unlocked. \\
\hline
Reset page counter &
Reset page use counter
based on page validation bit. \\
\hline
Acknowledge interrupts
\newline
(enhanced) &
Acknowledges all in-service and pending
interrupts in all processors to
avoid blocked interrupts after recovery. \\
\hline
Clear ``running'' VCPU flag &
Clear the VCPU flag that indicates
the VCPU is currently running before
rescheduling the VCPU. \\
\hline
\end{tabular}
 }
\end{table}

\begin{figure}
\centerline{\includegraphics[width=1.1in]{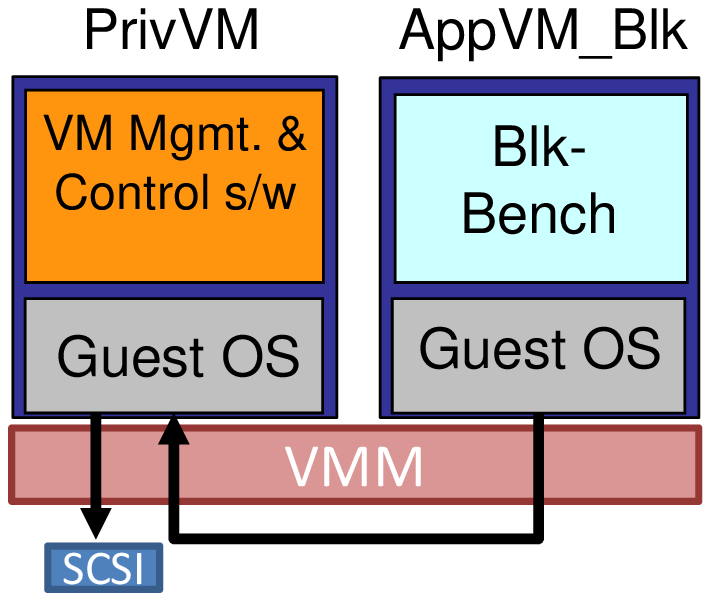}}
\caption{System configuration for 1AppVM.
AppVM\_Blk accesses block device (disk) through the PrivVM.
}
\label{fig:1appvm}

\vspace*{5ex}

\centerline{\includegraphics[width=2.2in]{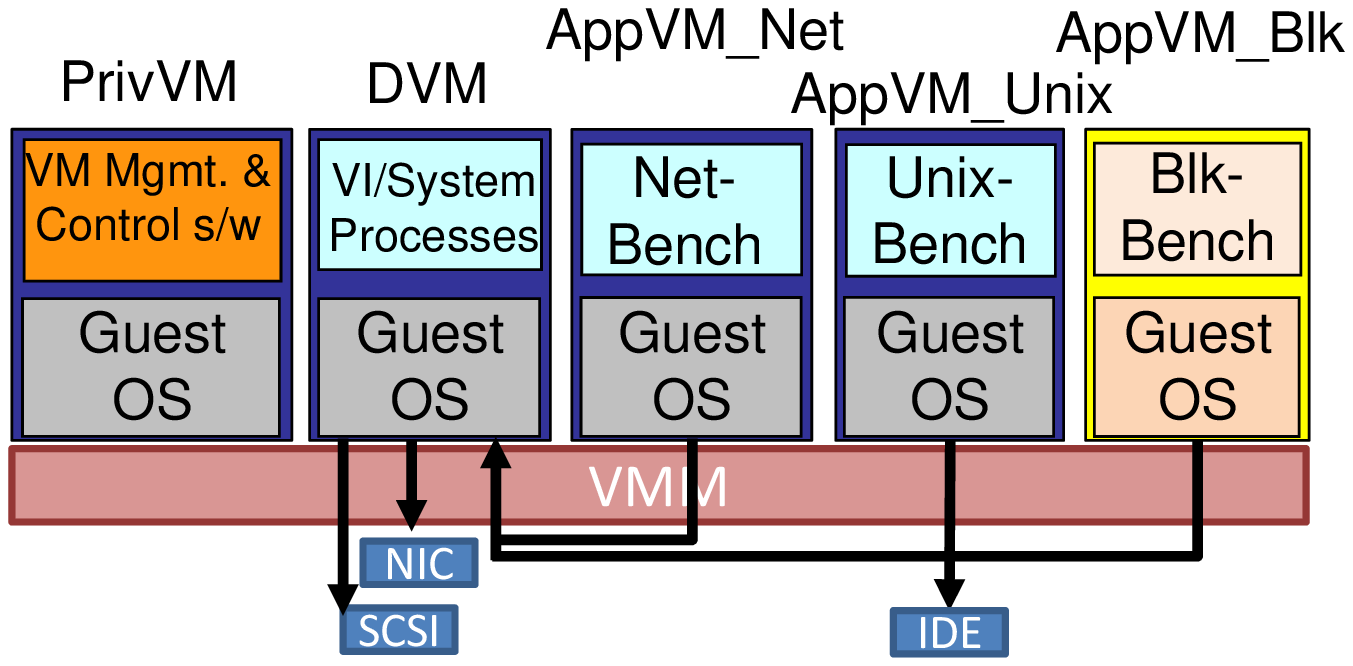}}
\caption{System configuration for 3AppVM.
AppVM\_Blk created after recovery.
AppVM\_Net and AppVM\_Blk accesses I/O devices through the DVM.
AppVM\_Unix accesses block device directly.
}
\label{fig:3appvm}

\vspace*{5ex}

\centerline{\includegraphics[width=2.4in]{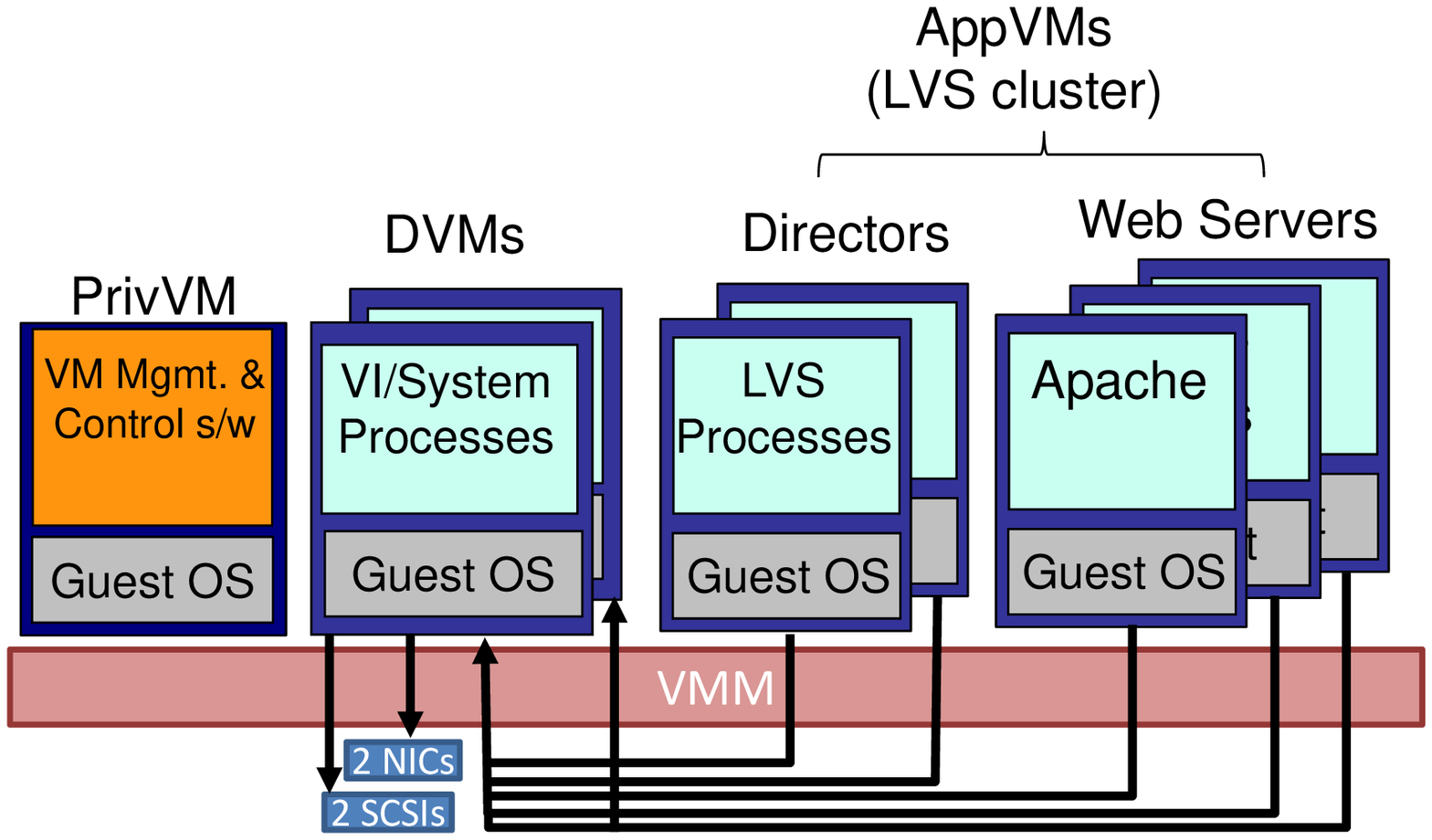}}
\caption{System configuration for 5AppVM.
AppVMs access I/O devices through DVMs.
Each AppVM has access to two disks, hosted on separate DVMs,
which the AppVM uses to form a single RAID level-1 block device.
Each AppVM has access to two NICs, with only one NIC
active at a time.
}
\label{fig:5appvm}
\end{figure}

We used software-implemented fault injection to
introduce errors into CPU registers when the VMM is executing.
The goal of the injection was to cause
arbitrary failures in the VMM and
evaluate the effectiveness of different recovery mechanisms.
Two system setups were used:
1AppVM and 3AppVM.
Details regarding these setups are presented in Section\ \ref{sec:exp}.
The 1AppVM setup (Figure \ref{fig:1appvm}),
with a single AppVM (AppVM\_Blk) running
a disk I/O (block) benchmark,
was used to quickly identify major shortcomings with the recovery
mechanisms.
The more complex 3AppVM setup (Figure \ref{fig:3appvm})
was used to
further stress the recovery mechanisms, once the majority of
sources of failed recoveries had been fixed.
The three AppVMs run different workloads.
Two of the three AppVMs (AppVM\_Net and AppVM\_Unix)
are booted when the entire system is booted.
The third AppVM (AppVM\_Blk) is booted after VMM recovery,
as a check of whether the recovered
virtualized system maintains the ability
to create new VMs.

\begin{table}
\tbl{
Injection outcomes.
\label{table:failoutcome}}{
\begin{tabular}{|l|p{7cm}|}
\hline
\multicolumn{1}{|c|}{Outcome} & \multicolumn{1}{c|}{Description} \\
\hhline{|=|=|}
Detected VMM failure &
\parbox{1cm}{Crash:} VMM panics due to
unrecoverable exceptions \newline
\parbox{1cm}{Hang:} VMM no longer makes
observable progress \\
\hline
Silent failure &
Undetected failure:
No VMM crash/hang detected but
applications in one or more VMs fail to complete successfully \\
\hline
Non-manifested &
No errors observed \\
\hline
\end{tabular}
 }
\end{table}

Table \ref{table:failoutcome} summarizes the possible outcomes
from an injected fault (an injection run).
Only detected VMM failures lead to VMM recoveries.
With the 1AppVM setup, a recovery is
considered successful if the
benchmark in AppVM\_Blk completes correctly.
With the 3AppVM setup,
following the explanation in Section \ref{sec:vmmfail},
a recovery is considered successful
if AppVM\_Net and/or AppVM\_Unix complete their
benchmarks correctly, and AppVM\_Blk
is able to boot and run its benchmark to completion without errors.
Silent failures,
discussed in Section \ref{sec:eval},
do not trigger VMM recovery
and are thus excluded from further discussion in this section.

The incremental enhancement of ReHype, discussed in this section,
is based on a sequence of fault injection campaigns,
with each successive campaign conducted on
a further enhanced version of ReHype.
As the scheme is improved, recovery failures become less
frequent, and more injections are needed per campaign in
order to identify the most important remaining cause
of recovery failures.
Thus, the number of injections performed in the different
campaigns progressively increases
from 300 to over 2800.

In the rest of this section,
each version of the recovery scheme is described,
and fault injection results are presented.
This is followed by an analysis
of the main cause of failed recoveries,
motivating the next version of the recovery scheme.
At each step, only the problem that led to the plurality of failed
recoveries is analyzed and fixed.
Table \ref{table:overallres} summarizes the rate of
successful recoveries with the basic ReHype scheme
and the various incremental improvements that were made.

\begin{table}
\tbl{
Recovery success rates out of detected VMM failures (crash/hang).
Target system: 1AppVM (Figure \ref{fig:1appvm}).
\label{table:overallres}}{
\begin{tabular}{|l|c|}
\hline
\multicolumn{1}{|c|}{Mechanism} & \multicolumn{1}{c|}{Successful Recovery Rate} \\
\hhline{|=|=|}
Basic & \phantom{0}5.6\% \\
\hline
+ NMI IPI & 17.6\% \\
\hline
+ Ack interrupts & 48.6\% \\
\hline
+ Hypercall retry & 62.6\% \\
\hline
+ FixSP+NMI ``ack'' & 77.0\% \\
\hline
+ Reinitialize locks & 95.8\% \\
\hline
\end{tabular}
 }
\end{table}

\textbf{Basic:}
As shown in Table \ref{table:overallres}, with the Basic
recovery scheme (Section \ref{sec:basicrec}),
the successful recovery rate is only 5.6\%.
A large fraction of recovery failures (44\%) occur
because the failure handler is unable to
initiate the VMM microreboot.
Normally, the failure handler relies on
interprocessor interrupts (IPIs) to force all processors
to save VM CPU state and halt execution
before microrebooting the VMM.
Microrebooting the VMM cannot proceed
until all processors execute the IPI handler.
Therefore, the failure handler is stuck
if a processor is unable to execute
the IPI handler due to a blocked IPI or memory corruption.

\textbf{NMI IPI:}
To get around the above problem, non-maskable interrupt (NMI)
IPIs are used.
In addition, a spin lock
protecting a structure used to set up an IPI function call
must be busted to prevent the failure handler from getting stuck.

Table \ref{table:overallres} shows an increase
in recovery success rate to 17.6\%
when these fixes are used.
Only 8.2\% of the failures
are now caused by an inability to initiate the
VMM microreboot.
The plurality of the remaining failures (45\%) are due to
interrupts from the block device not getting delivered to the PrivVM.
This causes the block device driver in the PrivVM
to time out, thus leading to the failure of
block requests from the AppVM.

The block device uses level-triggered interrupts.
For such interrupts,
the I/O controller blocks further interrupts until an
acknowledgment from the processor arrives.
If the VMM fails before acknowledging pending interrupts,
those level-triggered interrupts remain blocked after recovery.

\textbf{Acknowledge interrupts:}
To prevent level-triggered interrupts from being blocked,
the failure handler must acknowledge all pending interrupts
on all processors.
Pending interrupts can be classified as interrupts
waiting to be serviced by the VMM or interrupts
currently being serviced by the VMM (in-service).
Acknowledging in-service interrupts
can be easily accomplished
by executing the \textit{EOI} instruction
for every interrupt that has been delivered to the VMM.
It is necessary to perform
this operation
in the failure handler since information
about pending interrupts in the CPU
are cleared after a CPU reset during a VMM reboot.
A more complete (and more complex) approach
at acknowledging \textit{all} pending interrupts,
including interrupts waiting to be serviced,
is discussed in Section \ref{sec:enhadd}.

Table \ref{table:overallres} shows that when this
mechanism is added,
the successful recovery rate jumps to 48.6\%.
Of the remaining unsuccessful recoveries, 52.8\% are caused
by a crashed AppVM or PrivVM after recovery.
The crashes are caused by bad return values from hypercalls.
Since VMM failures can occur in the middle of a hypercall,
it is necessary to be able to transparently continue the
hypercall after recovery.
Without mechanisms to do this, after recovery, the VM
starts executing right after the hypercall, using whatever
is currently in the EAX register as the return value.

\textbf{Hypercall retry:}
The ability to restart a hypercall is already provided
in Xen.
The mechanism involves changing the program counter (PC)
of the VCPU to the address
of the instruction that invokes the hypercall.
For each VM,
the VMM determines whether a hypercall retry is needed
after the VMM microreboot, before loading the VM state.
Specifically,
for each VCPU,
the VMM checks if the VCPU's PC is within the VM's hypercall page.
If so, the VMM updates the VCPU's PC. 
Arguments to the hypercall are already preserved in the VM VCPU state.

Table \ref{table:overallres} shows
that, with hypercall retry, the successful recovery rate is 62.6\%.
Out of the remaining unsuccessful recoveries, 41\%
are caused by the same symptom encountered and partially solved
with the Basic scheme ---
the inability
of the failure handler to initiate the
VMM microreboot.
With the improved recovery rate, the causes of this
symptom not previously resolved are now responsible for
the plurality of failed recoveries.

The experimental results show
two causes for the symptom above:
(1) NMI IPIs sent to the wrong destination CPU due to
stack pointer corruption
and
(2) NMIs are blocked due to the Xen
NMI-based watchdog hang detection.
Problem (1) occurs because a corrupted stack pointer is used to obtain
the CPUID of the currently running processor.
The obtained CPUID is incorrect and is, in turn,
used to create a CPU destination mask for the NMI IPI.
This mask can end up containing the sending processor as one of
the destination CPUs.
The result of this is that an IPI is incorrectly sent
to the sending processor.
This IPI is dropped and the sender waits forever for the completion
of the IPI handling.

Problem (2) is due to the fact that NMI delivery is blocked
if a CPU is in the middle of handling a previous NMI ---
an \textit{iret} instruction matching a previous NMI
has not been executed
\cite{Intel10}.
The Xen hang detector is based on periodic NMIs from
a watchdog timer.
If a hang is detected on a processor, that processor
immediately executes the panic handler and
never executes an \textit{iret} instruction.
This prevents the processor from getting an NMI IPI from
the boot processor to initiate recovery.

\textbf{FixSP+NMI ``ack'':}
Problem (1) above can be fixed by not relying on the
stack pointer to obtain the
CPUID during failure handling.
Instead, the CPUID can be obtained by first reading the
APICID from the CPU
and then converting the APICID to CPUID,
using an existing APICID to CPUID mapping structure
stored in the static data segment of Xen.
With this technique, the VMM has a chance to continue
with the recovery despite a corrupted stack pointer.
However, the corrupted stack pointer can cause
critical problems that are unrelated to the CPUID.
Specifically, the handler invoked when VMM failure
is detected must save VCPU registers
(located on the stack) into preserved VMM state.
A corrupted
stack pointer leads to saving the contents
of a random region in memory as the saved VCPU register values.
At a later point in time,
this can lead to execution at an arbitrary location
in memory, with VMM privilege, leading to a VMM crash.
Specifically,
when attempting to load the saved VCPU registers after recovery,
the VMM may try to restore a corrupted value as the
VCPU code segment register.
This may cause the VMM to continue executing with VMM
privilege using corrupted (incorrect) register values.

ReHype implements a solution to Problem (1) above
that avoids the deficiency described in the previous paragraph.
Specifically, the failure handler, invoked upon VMM failure,
sets the stack pointer to a ``safe'' value.
This can be done based on the observation that
the failure handler never returns, and therefore, the stack pointer
can be reset to any valid stack location.
The address of the bottom of the stack
is kept by Xen in a static data area.
The stack pointer is set to that value
minus sufficient space
for local variables used by the failure handler.

Problem (2) above is resolved by forcing the execution of \textit{iret} in
the failure handler.
The values at the top of the stack are set
so that the \textit{iret} instruction
returns back to the failure handler code.

With the two improvements above, the rate of
successful recoveries is 77.0\%.
The majority of the increase is due
to fixing the stack pointer.
Since hangs are responsible for only a small fraction (7.1\%)
of detected VMM failures, the impact
of fixing problem (2) on the overall recovery success rate is small.
However, with this fix, there was successful recovery from
all hangs detected in this set of inejction runs.

Out of the
remaining unsuccessful recoveries, 82.8\%
are due to spin locks being held after recovery.
Spin locks that are statically allocated are re-initialized
on boot, but locks that are on the heap are not.
This causes the VMM to hang immediately after recovery.

\textbf{Reinitialize locks:}
Re-initializing dynamically-allocated spin locks requires
tracking the allocation and de-allocation of these locks.
All locks that are still allocated upon recovery are
initialized to unlocked state.
This tracking of spin locks is the only extra work
that ReHype must perform during normal operation.
The associated performance overhead is negligible since
the allocation and de-allocation of spin locks is normally
done only as part of VM creation and destruction.
Furthermore,
there are only about 20 spin locks that are tracked per VM.

Locking mechanisms that are not spin locks
must also be re-initialized to their free states.
A key example of this are
the page lock bits used to protect
access to bookkeeping information of pages.
With the previous version of the recovery scheme,
not initializing these bits resulted in 10\%
of unsuccessful recoveries.

As shown in Table \ref{table:overallres},
re-initializing locks increases successful recovery rate to 95.8\%.
For the remaining recovery failures there is no
one dominant cause.

While the 1AppVM system setup
is useful for uncovering the main
problems with the Basic ReHype recovery,
it is very simple, thus potentially hiding
important additional problems.
To better stress the virtualized system,
the rest of the experiments in this section
use the 3AppVM setup.
The results with this setup are summarized in
Table\ \ref{table:2contres}.

As shown in Table \ref{table:2contres},
with the 3AppVM setup,
the reinitialize locks mechanism
results in a recovery success rate of 88.6\%.
Hence, there is a decrease in the success rate
compared to the 1AppVM setup.

\begin{table}
\tbl{Recovery success rates out of detected VMM failures (crash/hang).
Target system: 3AppVM (Figure \ref{fig:3appvm}).
\label{table:2contres}}{
\begin{tabular}{|c|c|}
\hline
Mechanisms & Successful Recovery Rate \\
\hhline{|=|=|}
Reinitialize locks & 88.6\% \\
\hline
+ Reset page counter & 92.2\% \\
\hline
\end{tabular}
 }
\end{table}

Out of the remaining recovery failures,
about 35\% are due to the VMM
hanging immediately after recovery.
This problem is caused by
a data inconsistency resulting from a VMM failure
while in the middle of handling a page table update hypercall.
This hypercall promotes an unused VM page frame into a
page table type by incrementing a page type use counter
and performing validity check on the page frame.
After the validity check, a validity bit is set to indicate that
the page can be used as a page table for the VM.
Inconsistency arises when a
VMM failure occurs before the validity
check is completed but after
the page type use counter has been incremented.
When the hypercall is retried after recovery, since
the page use counter is not zero and the validity
bit is not set, the VMM code assumes that validation is in progress
and waits by spinning.
Of course, there is no validation taking place, and the CPU is declared
hung by the hang detector.

\textbf{Reset page counter:}
To fix the above problem,
the VMM bootup code is modified to check
the consistency between the validity bit and page use counter.
If the page type use counter is non-zero but the validity
bit is not set, then the page type use counter is set to zero.

With the page counter fix employed, recovery success rate
improves to 92.2\%.
The remaining causes of failed recoveries vary widely
and are discussed further in Section \ref{sec:eval}.
 \section{Experimental Setup}
\label{sec:exp}

This section presents the experimental setups used
to evaluate and validate
ReHype and the RVI as a whole.
Specifically, this section discusses
details of the different system configurations
used for stressing the recovery mechanisms
of the RVI,
the different workloads that are used,
the fault injection campaigns and fault types,
the fault injection outcomes, and
the failure detection mechanisms.

\subsection{System Configurations}
\label{sec:expsysconf}

The physical machines used for running
experiments are equipped with 8GB of memory
and dual quad-core Intel processors (Nehalem or Core\ 2).
In general, the virtualized system under evaluation
is comprised of
the Xen VMM, augmented with ReHype,
hosting the PrivVM, possibly a DVM
(depending on the campaign),
and one or more AppVMs.
All experiments in this work make use of one of three basic
system configurations: \mbox{1AppVM}, 3AppVM, and 5AppVM
(Figures \ref{fig:1appvm}-\ref{fig:5appvm}).
In one set of campaigns (Section\ \ref{sec:evalfv}),
the AppVMs are FV VMs.
In all the rest, they are PV VMs.

With the 1AppVM configuration (Figure \ref{fig:1appvm}),
the VMM hosts two VMs:
a PrivVM (\mbox{Domain 0}) and a single AppVM (AppVM\_Blk).
AppVM\_Blk
runs the \textit{Blkbench} benchmark (Subsection\ \ref{sec:workloads}),
which continuously performs disk I/O (block) operations.
The PrivVM hosts the block backend for the AppVM\_Blk.
Each of the VMs consists of one virtual CPU (VCPU)
that is pinned to its own physical CPU (PCPU).

\begin{figure}
\centerline{\includegraphics[width=2.8in]{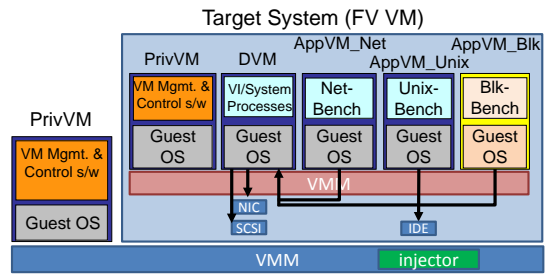}}
\caption{Target system is 3AppVM deployed inside a FV VM.
Injection is performed from the ``outer'' VMM.}
\label{fig:2contvmcreate}
\end{figure}

With 3AppVM (Figure \ref{fig:3appvm}),
there are three AppVMs:
AppVM\_Blk, AppVM\_Net, \mbox{AppVM\_Unix},
running benchmarks \textit{Blkbench}, \textit{Netbench},
and \textit{UnixBench}, respectively (Subsection\ \ref{sec:workloads}).
The PrivVM's root filesystem
is in memory and the PrivVM does not access any devices.
A separate Driver VM (DVM)
\cite{Fraser04}
hosts the backend drivers for AppVM\_Blk and AppVM\_Net.
AppVM\_Unix has direct access to an IDE controller,
and thus does not rely
on the DVM for any device access.
Each VM consists of one VCPU, which is pinned to
its own PCPU.
To check whether the recovered system maintains
its ability to create new VMs,
the system attempts to boot AppVM\_Blk
after a possible VMM recovery to run the
\textit{Blkbench} benchmark.

Much of the evaluation of ReHype is 
with the VMM hosting paravirtualized (PV) AppVMs.
The system must be able to host PV VMs since the
critical PrivVM and Driver VMs are PV VMs.
Furthermore, for the purpose of stressing ReHype,
PV VMs are a good choice
since the VMM is highly involved
in supporting
many typical VM operations, such as
page table updates, process scheduling, and timer operations.
The use of ReHype with the VMM hosting
fully-virtualized (FV) AppVMs is presented and evaluated in
Section \ref{sec:evalfv}.

For many of the experiments using the
1AppVM and 3AppVM configurations (except Sections
\ref{sec:validate}, \ref{sec:evalfv}, and \ref{sec:evallatency}),
the entire target system was run in a fully-virtualized (FV) VM.
Figure \ref{fig:2contvmcreate} shows this setup
when the target system is the 3AppVM configuration.
This setup simplified and sped up the fault injection campaigns
by facilitating the restart of the target system
and refresh its disk images after each injection run
to isolate the effects of faults injected in different runs
\cite{Le08,Le14}.
Since there is a potential that running the target system
in a VM may bias the results,
we have run experiments to validate the 3AppVM results
on bare hardware (Section\ \ref{sec:validate}).

The 5AppVM configuration (Figure \ref{fig:5appvm}),
consists of a PrivVM, two DVMs, and five AppVMs.
The five AppVMs together run a reliable Web service
workload (Subsection\ \ref{sec:workloads}).
Experiments using the 5AppVM configuration
run directly on bare hardware.
As with the 3AppVM configuration,
the PrivVM's root filesystem is in memory and the PrivVM
does not access any devices.
The AppVMs access their I/O devices
through the DVMs.
For seamless operation despite failure of a DVM,
for each AppVM, the root disk
is configured as a RAID Level-1 device, with each disk of the RAID
hosted on a separate DVM
\cite{Le11,Le13}.
The AppVM's network device is accessed through a single DVM
with the other DVM acting as a hot-spare.

\subsection{Workloads}
\label{sec:workloads}

For much of the evaluation of ReHype,
we use synthetic workloads,
consisting of three micro benchmarks:
\textit{Blkbench}, \textit{Netbench}, and \textit{UnixBench}.
\textit{Blkbench}
stresses the interface to the block device (disk) by
creating directories and creating, removing,
and copying 1MB files.
To ensure block activity,
this benchmark prevents caching of
block and filesystem data by the AppVM's OS.
When evaluating the results of each injection run,
an unsuccessful application completion
is recorded if:
(1)\ the application reports errors (failure of I/O operations),
and/or
(2)\ at the end of the run, the files and directories
created differ from a reference image.

\textit{Netbench} is a user-level \textit{ping} program that 
exercises the network interface.
It consists of two processes:
one running in an AppVM (the \textit{VM host}),
and another running on a separate
physical machine (the \textit{PM host}).
Every 1ms, the PM host transmits
a UDP packet to the VM host, which,
upon receiving this packet,
transmits UDP packet back to the PM host.
For each injection run,
if a VMM failure is detected,
an unsuccessful application completion
is recorded if:
(1)\ there is a interruption of more than
10s at any time during a run, and/or
(2)\ at any time after the recovery procedure completes,
there is a 1s interval
during which the rate of packet reception
at the PM host drops by more than 10\% compared to
the rate during normal operation.
If no VMM failure is detected,
an unsuccessful application completion is recorded
if, at any time during a run,
there is a 1s interval
during which the rate of packet reception
at the PM host drops by more than 10\% compared to
the rate during normal operation.
While these failure criteria cover
complicated failure modes,
based on our experience,
the effects of a failed VMM on \textit{Netbench}
is typically simple ---
in the vast majority of cases,
the PM host simply stops receiving packets
from the VM host.

Our \textit{UnixBench} is
a subset of the set of programs in UnixBench
\cite{UnixBench},
with minor modifications to
improve logging and failure detection.
The selected programs were chosen for their ability to stress
the VMM's handling of hypercalls such as
virtual memory management and process scheduling.
For each injection run,
an unsuccessful application completion
is recorded if:
(1)\ one or more programs in UnixBench terminate
prematurely due to failed system calls, and/or
(2)\ the resulting program output differs from a reference output.

To evaluate
the overall resiliency of our RVI, we use a workload
that is more representative of practical
deployments of virtualization:
cluster middleware that provides high availability,
hosting Web service.
The middleware is the
Linux Virtual Server (LVS)
\cite{Zhang03,Linux}, running
on a cluster of VMs (virtual cluster).
LVS is an open-source
load-balancing solution for
building highly-scalable and highly-available servers
using clusters of servers.
Out of the five AppVMs (virtual servers),
three run the Apache web server and two
act as a primary and backup load balancers (directors)
that forward client requests to the servers.
The Web servers are stressed by five
instances of the Apache \textit{ab} benchmark running on a remote host.
Each one of
two of the \textit{ab} instances sequentially sends 3770 HTTPS requests
for a static web page.
Each one of three
of the \textit{ab} instances sequentially sends 365 HTTPS requests
for a dynamically-generated Web page.
The Apache Web server executes the \textit{Blkbench}
program to generate a response for each request for a
dynamically generated web page.

The LVS cluster is considered to have failed
if it is unable to service Web requests.
However, LVS director and server failover
involves terminating client connections.
Any time a director or real server fails,
existing client connections through the director or with the real server
are terminated.
Furthermore,
in the case that a new director becomes the new primary director,
all connections must be terminated and re-formed
using the new primary director.
Hence, for each injection run,
an unsuccessful application completion is recorded if
a client experiences more than two connection timeouts
during the run.

\begin{table}
\tbl{
Fault types used in fault injection campaigns.
\label{table:injtype}}{
\begin{tabular}{|l|p{9cm}|}
\hline
\multicolumn{1}{|c|}{Fault Type} & \multicolumn{1}{c|}{Description} \\
\hhline{|=|=|}
Register & Flip random bit in a random general purpose register,
instruction pointer, stack pointer, or
system flags register {$^*$} \\
\hline
Code & Flip random bit in a random byte of an instruction \\
\hline
NOP & Replace random instrs with NOP \\
\hline
Destination & Flip random bit in destination operand of instruction \\
\hline
Source & Flip random bit in random source operand of instruction \\
\hline
Branch & Replace branch instructions with NOP \\
\hline
Loop & Reverse directions of loops \\
\hline
Pointer & Flip random bit in operand of memory access instructions \\
\hline
Interface & Use bad function arguments \\
\hline
\end{tabular}
 }
\begin{tabnote}
\tabnoteentry{$^*$}{
When the target system is run inside
a FV VM,
we do not inject into the
reserved bits and the VM-8086 bit
of the EFL register.
This is due to the limitation of performing
fault injection into a VM
as some hardware faults cannot be
accurately emulated
\cite{Le14}.
}
\end{tabnote}
\end{table}

\subsection{Fault Injection Campaigns and Fault Types}
\label{sec:inj}

The evaluation in this work is focused on
the \textit{recovery} mechanisms.
The fault injection campaigns used are designed
to ``stress'' the recovery mechanisms
in a variety of ways.
The goal is to first expose
problem areas and then evaluate the effectiveness
of the refined mechanisms to recover from a variety
of system corruptions.
We used the UCLA \textit{Gigan} fault injector
\cite{Le08,Hsu10,Le14}
to inject faults into different VI components.
Gigan can reside in the
VMM and inject many types of faults into the VMs and the VMM.
Injection into VMs can be done without any modifications
to the VMs.
Hence, with configurations where
the target system runs in an FV VM (Figure\ \ref{fig:2contvmcreate}),
Gigan runs in the outer VMM,
so injection does not require
any modifications (intrusion) of the target system.
Details of the fault injection campaigns and fault types
using the three different system configurations (1AppVM,
3AppVM, and 5AppVM) are discussed below.

A fault injection campaign consists
of many fault injection runs.
A single fault injection run that uses
the 1AppVM system configuration
consists of first booting the VMM along
with the PrivVM and AppVM\_Blk.
AppVM\_Blk begins running the blkbench benchmark and
a fault is injected into the VMM.
The injection campaign infrastructure
allows the target system sufficient time for
the VMM to recover and for the benchmark to complete.
If the benchmark does not complete, a timeout mechanism
identifies system failure.
At the end of each run,
fault injection logs and benchmark output are
retrieved and stored for analysis.

An injection run using the 3AppVM configuration
is similar to the 1AppVM configuration, except
that an injection is performed only after the VMM,
PrivVM, DVM, AppVM\_Net, and AppVM\_Unix have been booted and the two AppVMs
have started running their respective benchmarks.
Nine seconds after the two AppVMs begin running their benchmarks,
AppVM\_Blk is booted to run its own benchmark.
The injection run ends when all three AppVMs
complete their benchmark runs
or a timeout occurs.

For the 5AppVM configuration,
an injection run begins by booting
the VMM, PrivVM, two DVMs, and five AppVMs.
After the remote clients begin to generate requests,
a single fault is injected into one of the VI components.
Mechanisms are added to restart a node (AppVM)
the director deemed as failed.
If no nodes are restarted during a run,
one AppVM is randomly selected to be destroyed
and recreated
about 50 seconds after a fault is injected,
to ensure that, if recovery had occurred,
the VI is still correctly providing basic
functionality.

Three types of faults are used to evaluate ReHype:
random single bit-flips
in CPU registers during execution of VMM code,
random single bit-flips
in the VMM code segment (Code), and
software faults in the VMM (SW).
Bit-flips in CPU registers are used since most
transient hardware faults in CPU logic and memory
are likely to be manifested as erroneous
values in registers.
Furthermore, these faults can cause arbitrary corruptions in the
entire system.
Table \ref{table:injtype}
shows the different fault types
in more detail.
We refer to
the last seven rows of Table \ref{table:injtype}
collectively as software faults, since they
simulate typical programming errors
\cite{Ng01}.

An injection is triggered after a random time period
between 500ms to 6.5s after the AppVMs begin running their benchmarks.
For Code and SW faults,
a breakpoint 
is used to trigger an injection.
We used the fault injection tool in
\cite{Swift05a}
to generate a list of injection targets.
The breakpoint is set on the target VMM instruction
after the designated time
has elapsed.
To increase the activation rate
of Code and SW faults,
we used the Xenoprof
\cite{Menon05}
sampling profiler to identify
the most frequently executed functions
in the VMM,
considering only the instructions in those functions
as possible targets for fault injection.
In each run, a breakpoint is set
on a randomly selected CPU of the target system.
For register injection,
to ensure that the injection occurs only when the VMM is executing,
a fault is only injected after the designated time has elapsed and
0 to 20,000 VMM instructions, chosen at random, have been executed.

To evaluate the
resiliency of our RVI as a whole,
we inject faults into CPU registers
while the CPU is executing:
1) the VMM,
2) the DVM (user and kernel-level), and
3) the PrivVM (user and kernel-level).
The parameters of register injection
are similar to what was described above.
We only inject into CPU registers in these experiments
as this type of fault can not only be used to
emulate the effects of some software faults,
but based on our evaluation of ReHype
(see Section \ref{sec:eval}),
can be equally stressful on the recovery mechanisms
as Code and SW faults.

\subsection{Fault Injection Outcomes}

Table \ref{table:failoutcome} summarizes
the possible outcomes of each injection run
when evaluating ReHype (the target is only the VMM).
The outcome of each injection run
when evaluating the entire RVI
is classified in a similar way:
\textit{detected} (VMM/PrivVM/DVM crash or hang),
\textit{silent} (undetected failure),
or \textit{non-manifested}.
In general,
a crash occurs when one or more VI components panic
due to unrecoverable exceptions.
A hang occurs when a VI component
does not perform its expected operation
in a timely manner.
A silent failure
occurs when no VMM/DVM/PrivVM hang or crash
is detected but:
(1) the VI fails to host or create new AppVMs
and/or
(2) the applications (workload)
in one or more AppVMs fail to complete successfully.
What constitutes unsuccessful completion is application
specific, as discussed in Subsection\ \ref{sec:workloads}.
Non-manifested means that no errors are observed.

\subsection{Failure Detection}

We rely on simple techniques to detect when a VI
component has crashed or hanged.
In particular,
a crash is detected when the VMM
or the kernel of the PrivVM or DVM invokes
the panic handler
due to unrecoverable exceptions.
VMM hangs are detected using a watchdog mechanism built into Xen.
Specifically, Xen maintains a watchdog counter that is supposed to
be incremented by a normal timer event every 100ms.
A watchdog NMI is generated every 100ms
of unhalted CPU cycles.
If the watchdog NMI handler detects that the watchdog counter
has not been incremented for 300ms, the system is declared hung.
The detection of hangs of the DVM and PrivVM
relies on mechanisms we have added to the VMM and
are described in Subsections \ref{sec:umdvm} and \ref{sec:umprivvm},
respectively.
 \section{Impact on ReHype of VMM Support for PrivVM Recovery}
\label{sec:ftvi}

While the ability to recover from a failed VMM
is critical for VI resiliency, there is also
a need to handle failures of the
other VI components -- PrivVM and DVM. 
Hence, our RVI includes, in addition to ReHype,
mechanisms for tolerating failures
of DVMs \cite{Le09,Le13} and the PrivVM \cite{Le12}.

The experimental results reported in this paper
up to this point were obtained using a system that
included only ReHype and the DVM recovery mechanisms,
with the DVM recovery mechanisms disabled.
Our mechanism for PrivVM recovery required
modifications to the VMM \cite{Le12}.
In addition, we modified the mechanism
for logging information from the VMM.
The focus of this section is on assessing
the extent to which these VMM modifications (discussed
further below), affected the effectiveness of ReHype.
We refer to the RVI version with only ReHype and DVM
recovery mechanisms as RVI\subscript{HD}, and
with all mechanisms, including
the PrivVM recovery, as RVI\subscript{HDP}.

VMM modifications for PrivVM recovery
involved the addition of
about 365 lines of code
and the use of about 128MB of
VMM memory to store the PrivVM's kernel and filesystem images
(see Subsection \ref{sec:umprivvm} for details).
Additional modifications were made to reduce the
intrusion of logging information from the VMM
at the point of failure for post-mortem analysis.
Specifically, results of experiments with ReHype suggested that
some failed recoveries may be caused by logging operations
accessing corrupted state in the failed VMM.
Hence, modifications included disabling
outputs by the VMM to the serial console
as well as the use of per-CPU memory buffers,
instead of a single shared buffer,
to store outputs from the VMM's crash handler,
thus eliminating the need for locks to maintain write ordering.

Modifications to the VMM require revalidating ReHype
since such modifications can alter the alignment and location of
critical data structures in the VMM,
which can affect the outcome of faults.
As an example, consider an unrecoverable VMM failure due
to a fault in the stack pointer leading to
corruption of a critical VMM data structure located
near the VMM's stack.
This failure may not occur if that same critical
data structure
is allocated in a different area of memory.
Since the focus is on ReHype,
both the DVM and PrivVM recovery mechanisms were disabled
for this revalidation.

\begin{table}
\tbl{
Comparison of recovery success rates of ReHype
in RVI\subscript{HDP} vs. RVI\subscript{HD}.
RVI\subscript{HD} results shown in parenthesis.
\label{table:cmpoldnew}}{
\begin{tabular}{|c|l|c|}
\hline
Configuration & \multicolumn{1}{|c|}{Mechanisms} &
\vtop{\hbox{\strut Successful}\hbox{\strut Recovery Rate}} \\
\hhline{|=|=|=|}
1AppVM & Reinitialize locks & 93.3\% (95.8\%) \\
\hline
3AppVM & Reinitialize locks & 83.0\% (88.6\%) \\
\hline
3AppVM & + Reset page counter & 88.1\% (92.2\%) \\
\hline
\end{tabular}
 }
\end{table}

Table \ref{table:cmpoldnew} shows a comparison of
successful recovery rates with ReHype in
RVI\subscript{HDP} vs. RVI\subscript{HD}
(RVI\subscript{HD} results from Section\ \ref{sec:improve}).
The trends with the two RVI versions are similar:
in both versions there is a decrease
in the success rates when going from the simple 1AppVM
configuration to the more complex 3AppVM configuration
and an improvement in success rates
when the last recovery enhancement
discussed so far
(Reset page counter) is applied.
However, with all configurations,
the success rate in RVI\subscript{HDP} is lower.
This difference is likely to be related to
differences in exactly where
critical data structures,
such as descriptor tables and page directories/tables,
are located in memory.
Corruption of these data structures
will often lead to failure during
a VMM microreboot.
Specifically, close to 19\% of failed recoveries
with RVI\subscript{HDP}
were caused by a combination of fatal page
faults during the reboot of the new VMM
and corruption of segment descriptor tables
leading to the crash of the entire target system (FV VM)
by the outer VMM hosting the target system
(Subsection\ \ref{sec:expsysconf}).
These particular types of failures,
while possible,
did not occur with RVI\subscript{HD}.

RVI\subscript{HDP} is used
in the rest of this paper.

 \section{Enhancements to ReHype with Respect to Interrupts and VM Control}
\label{sec:enhadd}

In the process of working with ReHype and analyzing
experimental results,
we identified two additional simple enhancements
that increase ReHype's and our RVI's overall effectiveness.
The first enhancement improves the handling
of interrupts that are pending, but not yet
being serviced, at the time of VMM failure.
This enhancement reduces the number of cases where,
after VMM recovery, a single AppVM fails.
The second enhancement is a small modification of the
way the VMM handles the \textit{VM pause} hypercall,
to reduce VMM hangs after recovery.
This section is focused on these enhancements.

As discussed in Section\ \ref{sec:improve},
a significant enhancement of ReHype is a
mechanism that, when a VMM failure is detected,
acknowledges all in-service interrupts, thus
significantly reducing failures caused by
level-triggered interrupts blocking future
interrupts from the same device.
This \textit{acknowledge interrupts} mechanism
only acknowledges interrupts that are currently being
\textit{serviced} by the CPU.
In fact, a CPU can \textit{only} acknowledge an interrupt
that is currently being serviced by that CPU.
However, there may also be interrupts that are pending,
waiting to be serviced by the CPU.
Such interrupts cannot be acknowledged by
the mechanism described in Section\ \ref{sec:improve},
and, for level-triggered interrupts, block future
interrupts from that same device.

To resolve the above problem,
the VMM failure handler is modified to service
and acknowledge \textit{all} pending interrupts.
This must be done prior to rebooting the VMM, since
booting the VMM resets the CPU,
causing all information regarding pending
interrupts will be lost.
The enhanced \textit{acknowledge interrupts} mechanism
first masks all interrupts at the I/O controller,
preventing new interrupts from being sent to the CPUs.
The CPUs then install a new interrupt descriptor
table that contains dummy interrupt handlers.
These handlers simply acknowledge the interrupt and return.
Interrupts are then enabled on each CPU in order to flush
all pending interrupts and allow them to be acknowledged.
Recovery of the VMM resumes when there are no more pending
interrupts (checked by reading
the Interrupt Request Register on the CPU).

\begin{table}
\tbl{
The impact of the enhanced 
\textit{acknowledge interrupts} mechanism.
System configuration: 3AppVM with
RVI\subscript{HDP}.
\label{table:2contftvihdpres}}{
\begin{tabular}{|l|c|c|}
\hline
\multicolumn{1}{|c|}{Mechanisms} &
\parbox[t]{3.5cm}{\centering Successful Recovery Rate \\
   \textit{as defined in Section \ref{sec:vmmfail}} } &
\parbox[t]{3.5cm}{\centering Rate of Recoveries \\
   with no AppVM failure} \\
\hhline{|=|=|=|}
Reset page counter & 88.1\% & 63.7\% \\
\hline
+ Enhanced \textit{ack interrupts} & 89.1\% & 73.0\% \\
\hline
\end{tabular}
 }
\end{table}

Table \ref{table:2contftvihdpres} shows the impact of
the enhanced \textit{acknowledge interrupts} mechanism.
The impact on the successful recovery rate is small,
when ``successful recovery'' is defined as
in Section\ \ref{sec:vmmfail}.
However, the rate of recoveries in which
not even a single AppVM fails increases significantly.
This is due to the fact that in many of the successful
recoveries in which a single AppVM failed,
that single AppVM was AppVM\_Net and the failure
was caused by blocked interrupts from the network
interface card.

The evaluation of the resiliency of our complete RVI
(Section \ref{sec:rvs}) motivated another
enhancement to ReHype.
Specifically, in some cases, a single fault can cause
both the VMM and a DVM to fail.
In such cases, the DVM failure may be detected
before the VMM failure is detected, resulting
in VMM recovery being initiated in the middle
of the DVM recovery.
We found that, in many such cases, the VMM fails
immediately after the VMM recovery process completes
and DVM failure handling resumes.
As explained below, this problem is due to
inconsistency among parts of
the VM state maintained by the VMM, and can be easily avoided.

The VMM maintains, for each VM, a \textit{pause counter}.
If that counter is non-zero, the VM cannot be scheduled to run.
The VMM maintains, for each VCPU, a \textit{running}
flag that indicates whether the VCPU is currently executing.
This flag is set when the VCPU is scheduled to run
and cleared when the VCPU is descheduled.
The inconsistency mentioned above is between the \textit{pause counter}
of a VM and the \textit{running} flags of 
the VM's VCPUs.

With our mechanism for recovery from DVM failure
(Subsection \ref{sec:umdvm}),
the first step is for the PrivVM
to pause the failed DVM, using the \textit{VM pause} hypercall.
When the \textit{VM pause} hypercall is invoked,
the VMM increments the VM's \textit{pause counter}
and sends an IPI to stop any of the VM's VCPUs
that are currently running.
The VMM then waits until all the VM's VCPUs
have been descheduled, by checking, for each of the VCPUs,
if the \textit{running} flag has been cleared.
The problem can manifest if VMM failure is detected after
the pause counter is incremented
and before one of the VM's VCPUs has been scheduled out.
Thus, the \textit{running} flag of that VCPU is not cleared.
After recovery, the \textit{VM pause} hypercall is
retried and ends up waiting forever for the
\textit{running} flag to be cleared.
The solution to this problem is straightforward:
upon VMM microreboot,
initialize the relevant part of the VMM state
to safe values (Subsection\ \ref{sec:recappr})
before any VCPUs are scheduled.
Specifically, all the \textit{running} flags
of all the VCPUs are cleared.
 \section{Analysis of ReHype}
\label{sec:eval} 

This section analyzes fault injection results for the final
version of ReHype,
with all the enhancements
from Sections \ref{sec:improve}, \ref{sec:ftvi},
and \ref{sec:enhadd}.
Subsection\ \ref{sec:evalrecfail}
is focused on the recovery success rates and
causes of VMM recovery failures
with different fault types.
Subsection\ \ref{sec:evalsilent}
is focused on the causes of silent VMM failures.

\subsection{Recovery Effectiveness}
\label{sec:evalrecfail}

While the experimental results discussed
in the previous sections were used to guide
enhancements of ReHype, this subsection
presents an evaluation of the effectiveness
of the final scheme.
This evaluation is based not only on
faults in CPU registers, but also
software faults and random single-bit VMM code corruption
(see Subsection \ref{sec:inj}).
In addition, the main causes of recovery failures
are discussed.

\begin{table}
\tbl{
Recovery success rates of ReHype
across different fault types:
register (Reg), software (SW),
and VMM code bit flips (Code).
Success rates shown with 95\% confidence intervals.
Target system: 3AppVM.
\label{table:2contftvihdpsfres}}{
\begin{tabular}{|c|c|c|}
\hline
\parbox[c][0.7cm][c]{2cm}{\centering Fault Type} &
\parbox[c][0.7cm][c]{3.5cm}{\centering Successful Recovery Rate \\
   \textit{as defined in Section \ref{sec:vmmfail}} } &
\parbox[c][0.7cm][c]{3.5cm}{\centering Rate of Recoveries \\
   with no AppVM failure} \\
\hhline{|=|=|=|}
Reg & 89.1\% $\pm$ 4 & 73.0\% $\pm$ 4 \\
\hline
SW & 88.1\% $\pm$ 6 & 67.7\% $\pm$ 8 \\
\hline
Code & 88.6\% $\pm$ 4 & 72.0\% $\pm$ 8 \\
\hline
\end{tabular}
 }
\end{table}

Table \ref{table:2contftvihdpsfres} shows the recovery
success rates of ReHype when deployed
on the \mbox{3AppVM} configuration (Subsection \ref{sec:expsysconf})
with different types of faults.
These results indicate that the
effectiveness of ReHype
is similar across the different fault types.

Overall, recovery using
ReHype failed in about 12\% of detected failures.
Based on the discussion in
Section \ref{sec:vmmfail},
failed VMM recoveries can be classified into
four categories:
(i) unsuccessful VMM reboot,
(ii) successful VMM reboot but all AppVMs fail
to complete their benchmarks successfully,
(iii) successful VMM reboot and benchmark in AppVM\_Blk
completes successfully but benchmarks in both AppVM\_Net
and \mbox{AppVM\_Unix}
fail to complete successfully, and
(iv) successful VMM reboot
and successful completion of benchmarks
in AppVM\_Net and/or AppVM\_Unix
but benchmark in AppVM\_Blk fail to complete successfully.

Across the three fault types,
between 1/3 and 2/3 of recovery failures are in
category (i) above.
The majority of these failures
are caused by triple fault exceptions generated
during the execution of the VMM failure handler, triggering
a hardware system reset.
A triple fault exception is generated if an exception
is triggered while trying to invoke the double fault handler.
A double fault exception is generated if an exception 
is triggered while
trying to invoke an exception handler.
The inability to invoke an exception handler
is generally due to the corruption of the interrupt descriptor
table or memory address mapping data structures,
such as page tables or the global descriptor table.
In our experiments, such corruption
occur most frequently when the injected fault affects
the stack pointer register.
In fact, nearly half of failures in category (i) for register
injection is due to stack pointer corruption
leading to triple fault exceptions.

Other recovery failures
in category (i) are caused by various VMM corruptions and
inconsistencies including:
(1) corruption of VM's VCPU registers, causing the new VMM to crash
after recovery when attempting to schedule the VCPU;
(2) corruption of the timer heap, which leads to a page fault
in the VMM when the old timer heap is walked to restore
timer events; and
(3) page table corruption, causing
the new VMM to page fault early in the boot code.

Recovery failures in categories (ii)-(iv) are generally caused by
multiple problems that may appear to be independent
but actually manifest due to a single fault.
These problems are typically instances of:
(a) VM kernel panics due
to error return values from hypercalls or
VM state corruption,
(b) VM I/O requests/replies and timer events not handled
due to the loss of virtual interrupts, and
(c) the VMM completes the recovery process but latent
corrupted state causes it to fail repeatedly following recovery.
In some cases, a single problem
affects multiple VMs and in other cases
different problems affect different VMs.
In all cases, more than one AppVM fails
to execute its benchmark successfully.
For AppVM\_Blk, this occurs most often
because, after recovery, the VMM is unable
to successfully create a new VM.

A representative example of how a hypercall can fail
after recovery and eventually result in a VM failure is
related to the hypercall retry mechanism (Section \ref{sec:improve}).
Specifically, it relates to the fact that
a hypercall that would have succeeded in normal
operation, fails when it is retried after
a VMM microreboot (the hypercall is not idempotent).
One such hypercall is used by a VM to unmap
a page that is shared with another VM (the page owner).
This informs the VMM that the caller is no longer
using the page and the appropriate VMM bookkeeping
(update the grant table) should be done.
If the VMM fails after the hypercall
has removed the mapping from the caller VM's page table,
the retry of the hypercall fails
since the hypercall expects the page to still be mapped.
This prevents the grant table from being properly updated.
At a later point in time, the VM that owns the page
fails (panics) when it tries to, again, share the page
with another VM.

\subsection{Silent Failures}
\label{sec:evalsilent}

\begin{table}
\tbl{
The impact of silent failures --
percentages of manifested faults
that result in silent failures.
``System \mbox{failures}'' are defined in
Section \ref{sec:vmmfail}.
Target system: \mbox{3AppVM}.
\label{table:2contftvihdpsilentres}}
{\begin{tabular}{|c|c|c|}
\hline
\multirow{2}{*}{Fault Type} &
   \multicolumn{2}{c|}{Silent Failures / Manifested} \\
\cline{2-3}
& 1 AppVM Failure & System Failure \\
\hhline{|=|=|=|}
Reg & 8.1\%  & 10.9\% \\
\hline
SW & 0.6\% & 35.0\%  \\
\hline
Code & 0.9\% & 22.5\% \\
\hline
\end{tabular}
 }
\end{table}

Faults during VMM execution
can lead to failures that are not detectable
by simple VMM crash and hang detectors.
Such failures are referred to as \textit{silent failures}.
A simple example of such a scenario is a fault
that causes the VMM to corrupt the states of multiple VMs,
leading to the subsequent failure of all of those VMs.
Table \ref{table:2contftvihdpsilentres}
shows the percentages of silent failures
out of manifested faults for the different
VMM fault types.
The single AppVM failures are mostly either \textit{Netbench}
or \textit{UnixBench} failing to complete successfully,
due to failed hypercalls or blocked/lost virtual interrupts.

Silent system failures can be classified
similarly to recovery failures
(Subsection\ \ref{sec:evalrecfail})
into four categories:
(i) failure of the entire system
(i.e., hardware reset),
(ii) all AppVMs fail to complete their benchmarks successfully,
(iii) benchmark in AppVM\_Blk completes successfully but
benchmarks in both AppVM\_Net and AppVM\_Unix
fail to complete successfully, and
(iv) benchmarks in AppVM\_Net and/or AppVM\_Unix complete successfully
but benchmark in AppVM\_Blk fails to complete successfully.

The main causes of silent system failures for each
of the categories are similar to those
for recovery failures.
Specifically, for category (i), the main cause
of failures are triple faults generated
before VMM failures are detected.
For category (ii)-(iv),
there are two main causes of silent system failures:
(a) VM kernel panics due to error return values
from hypercalls or VM state corruption,
and
(b) VM I/O requests/replies and timer events
not handled due to the loss of virtual interrupts.

 \section{Validation of ReHype on Bare Hardware}
\label{sec:validate}

As explained in Section \ref{sec:exp},
the experimental evaluation of ReHype discussed
so far was performed with the target system
running in an FV VM (Figure\ \ref{fig:2contvmcreate}).
Since any deployment of ReHype in a production
environment would be on bare hardware,
it is important to determine whether
the effectiveness of ReHype on bare hardware is
significantly different.
This determination is the focus of this section.

\begin{table}
\tbl{
Recovery success rates of ReHype
\textit{on bare hardware}
across different fault types.
Success rates shown with 95\% confidence intervals.
Target system: 3AppVM.
Success rates with the same system running in an FV VM
(Table\ \ref{table:2contftvihdpsfres})
shown in parenthesis.
\label{table:resbare}}{
\begin{tabular}{|c|c|c|}
\hline
\parbox[c][0.7cm][c]{2cm}{\centering Fault Type} &
\parbox[c][0.7cm][c]{3.5cm}{\centering Successful Recovery Rate \\
   \textit{as defined in Section \ref{sec:vmmfail}} } &
\parbox[c][0.7cm][c]{3.5cm}{\centering Rate of Recoveries \\
   with no AppVM failure} \\
\hhline{|=|=|=|}
Reg & 90.9\% $\pm$ 7 (89.1\%) & 69.2\% $\pm$ 14 (73.0\%) \\
\hline
SW & 87.6\% $\pm$ 2 (88.1\%) & 72.1\% $\pm$ 7\hspace{0.65em} (67.7\%) \\
\hline
Code & 91.0\% $\pm$ 3 (88.6\%) & 73.0\% $\pm$ 15 (72.0\%) \\
\hline
\end{tabular}
 }

\vspace*{5ex}

\tbl{
The impact of silent failures on
\textit{on bare hardware} --
percentages of manifested faults
that result in silent failures.
Target system: 3AppVM.
Values in parenthesis are results
from running target system inside a FV VM
(copied from Table \ref{table:2contftvihdpsilentres}).
\label{table:silentresbare}}{
\begin{tabular}{|c|c|c|}
\hline
\multirow{2}{*}{Fault Type} &
   \multicolumn{2}{c|}{Silent Failures / Manifested} \\
\cline{2-3}
& 1 AppVM Failure & System Failure \\
\hhline{|=|=|=|}
Reg & 5.8\% (8.1\%) & 12.6\% (10.9\%) \\
\hline
SW & 1.2\% (0.6\%) & 36.0\% (35.0\%)  \\
\hline
Code & 1.3\% (0.9\%) & 22.3\% (22.5\%) \\
\hline
\end{tabular}
 }
\end{table}

A first attempt to deploy ReHype on bare hardware
failed with the VMM hanging during every microreboot.
As part of the boot process,
the VMM normally performs low-level BIOS accesses
to gather information about the hard disks,
using Enhanced Disk Drive Services,
and information about the display, by accessing
the Extended Display Identification Data.
We determined that these operations cause
the hang during the VMM microreboot.
We believe that this may be due to a bug in the BIOS
that manifests when the system is not
fully reset prior to booting.
Interestingly,
in order to overcome buggy BIOS implementations,
the option to skip the BIOS probing of
devices is already provided
as command line arguments to the Xen VMM as well as
the Linux kernel.

Overcoming the problem discussed above required
a minor modification to the VMM so that it skips
the BIOS probing of devices during microreboot.
It should be noted that, if needed, it is
straightforward to add the ability to save
the information obtained during the initial boot
from the BIOS probing and restore this information
during a microreboot.

Table \ref{table:resbare} shows the recovery success rates
of ReHype with the 3AppVM configuration (Figure\ \ref{fig:3appvm}),
deployed on bare hardware.
The injection campaign was the same as the one performed
with the entire target system deployed in a VM
(Subsection\ \ref{sec:evalrecfail}).
Since a fault injection run on bare hardware
takes much longer
to complete (roughly four times longer \cite{Le14}),
we injected fewer faults per campaign
than when the target system was deployed in a VM,
resulting in larger confidence intervals.
For ease of comparison, the table also includes
the results when the system is deployed in a VM
(Table\ \ref{table:2contftvihdpsfres}).
Table\ \ref{table:silentresbare} shows the percentages of
silent failures out of manifested VMM faults when
the system is deployed on bare hardware.
The results from the same measurements when the system
is deployed in a VM (Table\ \ref{table:2contftvihdpsilentres})
are also shown.

Based on Tables \ref{table:resbare} and \ref{table:silentresbare},
the impact of faults is essentially the same when the system is
deployed on bare hardware as when it is deployed in a VM.
This matches previous results with other target systems,
comparing the impact of
faults with the system deployed on bare
hardware vs. in a VM \cite{Le14}.
Furthermore, this result reinforces the validity of
various measurements presented in this paper with target
systems deployed in VMs.

 \section{Validation of ReHype with FV AppVMs}
\label{sec:evalfv}

As explained in Section \ref{sec:exp},
the experimental evaluation of ReHype discussed
so far was performed with all the VMs
using paravirtualized (PV) OS kernels.
However, in most deployments of virtualization,
utilizing hardware support for virtualization \cite{Uhlig05},
most of the AppVMs are fully virtualized (FV),
where the guest OS kernels are not modified
in order to run in VMs.
Hence, this section is focused on validating
ReHype with FV AppVMs.

We experimented with ReHype in a system with
FV AppVMs, where these AppVMs use
hardware-assisted paging \cite{AMD08}
and are configured with para-virtualized devices.
Processors with
hardware-assisted paging are widely available
and the mechanism is commonly used to reduce the overhead and
complexity of address mapping in virtualized systems.
When using device controllers without special support
for virtualization, para-virtualized devices
are sometimes used to allow sharing among VMs
while maximizing performance \cite{Koh09}.
For ReHype, there are significant advantages to using
hardware-assisted paging, as opposed to
the alternative of using shadow page tables,
and the use of para-virtualized devices,
as opposed to the alternative of fully-virtualized devices.
In both cases, these choices reduce VMM activity,
and thus the opportunities for
recovery failures due to inconsistencies 
among different parts of VMM state.

With the system setup described above, ReHype works
with FV AppVMs, without any modifications.
We evaluated ReHype with the 3AppVM configuration
(Figure\ \ref{fig:3appvm}) running on bare hardware,
with all three AppVM being FV VMs.
A difference in the setup compared to the
description in Subsection\ \ref{sec:expsysconf}
is that AppVM\_Unix accesses its block device through
the DVM rather than directly.

\begin{table}
\tbl{
Recovery success rates of ReHype
\textit{hosting FV AppVMs}
with the target system running on bare hardware
across different fault types.
Target system: 3AppVM.
Success rates shown with 95\% confidence intervals.
Success rates with the same system running on bare hardware
but hosting PV AppVMs (Table \ref{table:resbare})
shown in parenthesis.
\label{table:2contftvihdpfvres}}{
\begin{tabular}{|c|c|c|}
\hline
\parbox[c][0.7cm][c]{2cm}{\centering Fault Type} &
\parbox[c][0.7cm][c]{3.5cm}{\centering Successful Recovery Rate \\
   \textit{as defined in Section \ref{sec:vmmfail}} } &
\parbox[c][0.7cm][c]{3.5cm}{\centering Rate of Recoveries \\
   with no AppVM failure} \\
\hhline{|=|=|=|}
Reg & 88.2\% $\pm$ 6 (90.9\%) & 85.5\% $\pm$ 2\hspace{0.65em} (69.2\%) \\
\hline
SW & 82.9\% $\pm$ 6 (87.6\%) & 78.4\% $\pm$ 12 (72.1\%) \\
\hline
Code & 82.5\% $\pm$ 4 (91.0\%) & 79.6\% $\pm$ 5\hspace{0.65em} (73.0\%) \\
\hline
\end{tabular}
 }

\vspace*{5ex}

\tbl{
The impact of silent failures on a system
\textit{hosting FV AppVMs}
running on bare hardware --
percentages of manifested faults
that result in silent failures.
Target system: 3AppVM.
Values in parenthesis are results when the
system hosts PV AppVMs (Table \ref{table:silentresbare}).
\label{table:2contftvihdpfvsilentres}}{
\begin{tabular}{|c|c|c|}
\hline
\multirow{2}{*}{Fault Type} &
   \multicolumn{2}{c|}{Silent Failures / Manifested} \\
\cline{2-3}
& 1 AppVM Failure & System Failure \\
\hhline{|=|=|=|}
Reg & 7.9\% (5.8\%)  & 14.0\% (12.6\%) \\
\hline
SW & 2.1\% (1.2\%)  & 28.3\% (36.0\%) \\
\hline
Code & 3.5\% (1.3\%)  & 23.9\% (22.3\%) \\
\hline
\end{tabular}
 }
\end{table}

Table \ref{table:2contftvihdpfvres} shows the recovery success rates
of ReHype on the 3AppVM configuration (Figure\ \ref{fig:3appvm}),
deployed on bare hardware, with FV AppVMs.
For ease of comparison, the table also includes
the results when the system is hosts PV AppVMs
(Table\ \ref{table:resbare}).
Overall, the results with FV AppVMs are very similar
to those with PV AppVMs.
For SW and Code injections,
the successful recovery rate is a little lower with
FV AppVMs compared to with PV AppVMs.
This is mainly caused by
differences in the VMM instructions
targeted with the two configurations.
As described in subsection \ref{sec:inj},
the instruction targets are chosen
based on profiling the VMM and selecting
the instructions that belong
to functions most frequently executed. 
Since there are different activities
in the VMM when hosting FV and PV VMs,
the two configurations yield different
VMM execution profiles,
and thus, different sets of instruction targets.
We experimented with using the same set of
instruction targets for the two setups
and found that the recovery success rate were
essentially the same.

Table \ref{table:2contftvihdpfvsilentres} shows the percentages of
silent failures out of manifested VMM faults when
the system hosts FV AppVMs.
The results from the same measurements when the system
hosts PV AppVMS (Table\ \ref{table:silentresbare})
are also shown.
Here again, there are no statistically significant
difference between the two sets of results.
Overall, the results in this section demonstrate that
the effectiveness of ReHype is similar with
FV AppVMs and PV AppVMs.

 \section{Recovery Latency}
\label{sec:evallatency}

When VMM failure is detected,
Rehype pauses all the VMs on the system
and unpaused them only when the reboot of the
new VMM instance completes.
Hence, execution of all the applications running
on the VMs is blocked.
For some applications, the duration of this
service interruption is critical.
The focus of this section is on the analysis of
VMM recovery latency with ReHype and modifications
to minimize this latency.

We evaluate the recovery latency using the same
setup used in Section\ \ref{sec:validate}:
the 3AppVM configuration (Figure\ \ref{fig:3appvm})
running on bare hardware.
The impact of recovery on the \textit{Netbench} benchmark
(Subsection\ \ref{sec:workloads}) is used to
measure the recovery latency.
During normal operations, the ping inter-arrival time
of \textit{Netbench} is roughly 1.1ms.
During recovery, AppVM\_Net is paused, so the recovery
latency is measured by simply recording the change
in the ping inter-arrival time on the separate
physical machine.
For this measurement, the VMM crash handler is invoked
directly, thus excluding the detection latency.

\begin{table}
\tbl{
Breakdown of recovery latency using ReHype
with and without optimization.
\label{table:reclat}}{
\begin{tabular}{|p{4cm}|r|r|}
\hline
\multicolumn{1}{|c|}{Operations} &
\vtop{\hbox{\strut Time}\hbox{\strut (no opt.)}} &
\vtop{\hbox{\strut Time}\hbox{\strut (with opt.)}} \\
\hhline{|=|=|=|}
CPU initialization:
\newline
\begin{tabular}{@{}l@{ }p{3.5cm}}
- & Initialize and wait for all CPUs to come online
\end{tabular}
& 150ms & 150ms \\
\hline
Timer/hardware initialization:
\newline
\begin{tabular}{@{}l@{ }p{3.5cm}}
- & Initialize/calibrate platform timer,
TSC, I/O APIC, NMI watchdog, etc.
\end{tabular}
& 410ms & 310ms \\
\hline
Memory initialization: & 2330ms & 250ms \\
\begin{tabular}{@{}l@{ }p{3.5cm}}
- & Record page number of all
allocated pages in old heap
(Use to preserve content of old heap)
\end{tabular}
& 20ms & 20ms \\
\begin{tabular}{@{}l@{ }p{3.5cm}}
- & Restore and check consistency
of page frame entries
\end{tabular}
& 30ms & 30ms \\
\begin{tabular}{@{}l@{ }p{3.5cm}}
- &
Create and allocate free pages to
VMM's heap
\end{tabular}
& 200ms & 200ms \\
\begin{tabular}{@{}l@{ }p{3.5cm}}
- &
Scrub unallocated pages
\end{tabular}
& 2080ms & 0ms \\
\hline
Other & 5ms & 5ms \\
\hline
Total & 2895ms & 715ms \\
\hline
\end{tabular}
 }
\end{table}

Using the procedure described above, we measured
a recovery latency of 2895ms.
To obtain a breakdown of this latency into the main
steps involved in recovery, we added code to log
the timestamps of key events.
Table\ \ref{table:reclat} (middle column) shows the results.
The bulk of the recovery time is
spent by the VMM initializing hardware devices and creating
data structures associated with
the management of CPUs, memory,
platform timers, and interrupt controllers.
Specifically,
the majority of the recovery time is spent performing 
memory initialization operations,
with the bulk of that time spent
scrubbing (zeroing) all unallocated pages.
Scrubbing unallocated memory pages
is a security measure that prevents
the leaking of data among VMs.

To reduced the recovery latency, we identified
two time-consuming operations
that can be skipped on a VMM reboot:
the scrubbing of unallocated pages and
the check of whether the NMI watchdog mechanism works properly.
When a VMM microreboot is initiated,
unallocated pages are either already scrubbed
by the failed VMM
or are on a list of pages to be scrubbed.
Hence, assuming that the old VMM instance was
correctly performing memory scrubbing up until
VMM failure was detected, the page scrubbing step can
be skipped.

As explained in Sections \ref{sec:vmmfail} and \ref{sec:improve},
Xen includes a hang detection mechanism based on
periodic NMIs from a watchdog timer.
During reboot, there is a check of whether
the  NMI watchdog mechanism operates correctly.
This check involves counting
the number of NMI interrupts received by the
VMM within an interval of 100ms
and verifying that it is above a preset threshold.
Our second recovery latency optimization is to
skip this check during a VMM microreboot.
Together, as show in Table\ \ref{table:reclat},
the two optimizations reduce the recovery
latency from 2895ms to 715ms.

It should be noted that recovery time could be reduced
by modifying the VMM boot code to parallelize
some of the initialization operations.
For example, while waiting for CPUs
to come online, entries in the
old page frame table can be checked and restored.
Such optimizations would require a significant
engineering effort to refactor the VMM boot code.
 \section{A Resilient Virtualization Infrastructure}
\label{sec:rvs}

As discussed earlier, while ReHype is the key
mechanism for resilient virtualization,
it is also necessary to provide
resiliency to failures of other VI components:
the privileged VM (PrivVM) and driver VM (DVM).
This section presents resiliency mechanisms
for the PrivVM and DVM, and explains how the different
parts of our resilient VI (RVI) fit together, so that, as a whole,
the virtualized system can tolerate
any single VI component failure.
To facilitate the explanation,
a brief overview of key aspects of the Xen VI is presented in
Subsection\ \ref{sec:xenvi}.
Our RVI is presented in Subsection\ \ref{sec:ftviframe}.
Subsection\ \ref{sec:imp_complexity}
provides measures of the implementation complexity.

\subsection{Xen VI Overview}
\label{sec:xenvi}

In order to allow multiple VMs to share I/O devices,
Xen uses a mechanism called
the split device driver architecture
\cite{Fraser04,LeVasseur04,Microsoft}.
With this organization, a \textit{frontend driver}
resides in each AppVM sharing a device.
The actual device driver
as well as a \textit{backend driver}
reside in the DVM.
In each AppVM, I/O requests are forwarded
by the frontend driver to the backend driver,
which invokes the actual device driver.
In Xen
\cite{Barham03},
the frontend and backend drivers communicate
through a ring data structure in an area of memory
shared between the AppVM and DVM.

The privileged VM (PrivVM)
is used to perform system management operations, such as
creating, destroying, and checkpointing VMs.
The VMM does not permit these operations to be invoked
by any other VMs.
The functionality of the PrivVM is provided by
a combination of kernel modules
and user-level processes running in the PrivVM.
One user-level process, XenStored, provides
access to a dynamic database of system configuration
information, called XenStore.
XenStored also provides mechanisms
for VMs to be informed of changes to certain
configuration states by allowing VMs
to register \textit{watches} on those states in the XenStore.
A VM communicates with the XenStore through XenStored using
a shared ring data structure,
similar to the communication mechanism between a DVM and AppVM.

In most Xen deployments,
the PrivVM not only controls and manages other VMs,
but also serves as a DVM.
However, such a configuration makes
each of these components vulnerable to
failures of the other, and is thus a poor choice
for achieving resiliency.
Stock Xen allows the system to be configured so that
the DVM functionality is in a separate VM.

\begin{figure}
\centerline{\includegraphics[width=3.0in]{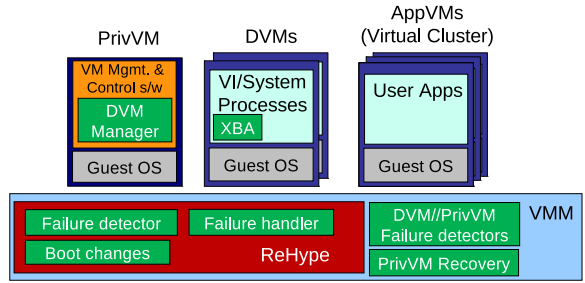}}
\caption{
A resilient virtualization infrastructure (RVI) based on Xen,
highlighting the main resiliency components.
}
\label{fig:sysvirtoverall}
\end{figure}

\subsection{Design and Implementation of a Resilient VI}
\label{sec:ftviframe}

Figure \ref{fig:sysvirtoverall} shows
the main components of our RVI
hosting a virtual cluster consisting of multiple
\hyphenation{AppVMs}
AppVMs.
The VI resiliency enhancements include:
ReHype for detecting and recovering from VMM failure,
two DVMs to enable uninterrupted access
to devices for the AppVMs, a DVM Manager for
controlling recovery from DVM failure,
DVM failure detectors for detecting and notifying
AppVMs and the DVM Manager of the DVM failure,
a XenStore Backup Agent (XBA),
and PrivVM failure detectors along with mechanisms
in the VMM and XBA to microreboot and restore
the state of a failed PrivVM.
The following subsections briefly explain how
these mechanisms provide resiliency to
DVM and PrivVM failures.

\subsubsection{DVM Recovery}
\label{sec:umdvm}

When a DVM fails,
applications accessing I/O devices through that DVM
are blocked.
DVM crashes are detected when
the crash handler in the DVM's kernel
makes a hypercall to the VMM, or
when the VMM responds to illegal DVM
activity by killing the DVM.
DVM hangs are detected when the DVM stops context
switching among processes or when the DVM
stops consuming requests on its shared rings
with AppVMs
\cite{Le09,Le13}.

If the DVM is microrebooted
and hardware devices are reset,
the duration of the interruption may be on the order of seconds
\cite{Le09}.
Such long interruptions can result in the failure of
the workload running in the AppVMs.
Therefore, unlike other VI components,
we do not rely on microreboot to recover from DVM failure.
Instead, recovery from a DVM failure involves
failing over to a redundant DVM with access to
separate hardware devices
\cite{Le11,Le13}.
However, microreboot must still be used to replace
the failed DVM, so that the
fault tolerance capabilities of the system are restored.

The PrivVM controls the process
of microrebooting the DVM, which includes:
pausing the failed DVM,
booting a new DVM instance, destroying the failed DVM,
and integrating the new DVM with existing VMs on the system.
The destruction of the failed DVM
and subsequent releasing of all its memory to the VMM
must be done after all the devices that the failed
DVM owns are re-initialized
by the newly booted DVM.
For systems without I/O MMUs
\cite{Ben-Yehuda06},
this particular ordering of events
can prevent ongoing DMA operations initiated by the failed
DVM from corrupting memory that has been released to the VMM.
The new DVM instance is re-integrated with existing VMs
by reforming the respective frontend-backend connections.
This is done transparently to the applications in the AppVM
by extending existing mechanisms in the frontend drivers
responsible for resuming and suspending devices
\cite{Le09,Le13}.

\subsubsection{PrivVM Recovery}
\label{sec:umprivvm}

The mechanisms used to detect crashes and hangs
of the PrivVM's kernel are similar to those
used for the DVM.
As described in Subsection \ref{sec:xenvi},
the PrivVM
hosts user-level processes that are essential
to the correct operations of the VI.
To detect the failure of these processes,
we use a user-level monitoring process, called \textit{hostmon},
that periodically checks for the existence of these processes
\cite{Le12}.
When hostmon detects the disappearance of one of these critical
processes, it uses a hypercall to
cause the VMM to crash the PrivVM and trigger
full PrivVM recovery.

Of the three VI components,
only the VMM has the privileges required to
recover a failed PrivVM.
Hence, the VMM is responsible for
releasing all the resources
of the failed PrivVM and booting the new PrivVM instance.
Since the PrivVM kernel and root file system may
be corrupted during PrivVM failure,
pristine PrivVM kernel and filesystem images must be
used for the new PrivVM.
The required pristine images are stored, compressed,
in the VMM address space, consuming approximately 128MB.

A key requirement for microrebooting the PrivVM is
to restore state in the PrivVM needed for managing
and controlling the system.
This state includes
the XenStore, stored as a file in the PrivVM,
and \textit{watches} in the XenStored process.
Since all the PrivVM state, including the file system,
is in memory,
failure of the PrivVM results in the complete loss of
its state.
Hence, to preserve the critical PrivVM state,
the XenStore and XenStored states are replicated.
To survive PrivVM failure,
the replicated states must be stored in a different
VI component.
While there are several alternatives,
the simplest choice is to maintain the replicated
state in one of the DVMs.
This DVM is referred to as DVM\_XS \cite{Le12}.

The backup copy of the critical PrivVM state
is maintained in the DVM\_XS
by a user-level process ---
the XenStore Backup Agent (XBA).
XenStore write requests,
watch registrations, and requests to 
start or end XenStore transactions
are forwarded by XenStored to the XBA
before performing the operations in the PrivVM.
The XBA performs all operations on a local
copy of the XenStore located on the
filesystem of the DVM\_XS.
After a microreboot, the new PrivVM acquires up-to-date
XenStore and XenStored states from the XBA.
If the DVM\_XS fails,
the XenStore and XenStored states
are transmitted from the PrivVM to the XBA
on the newly recovered DVM\_XS.

Simply maintaining a backup copy of critical
PrivVM state is insufficient
for the correct recovery of the PrivVM.
Failure of the PrivVM can lead to inconsistencies
in the state of the recovered
PrivVM and between the recovered PrivVM
and other VMs.
These inconsistencies can occur when the PrivVM
fails while in the middle of management operations
such as creating/destroying VMs or handling
XenStore requests from VMs.
To avoid these problems, we added mechanisms
to make VM management operations (VM create/destroy)
as well as XenStore request handling atomic \cite{Le12}.
Transactionalizing these operations
requires maintaining a log
that tracks the individual steps of each operation.
This log is kept safe on the DVM\_XS
and allows the recovery mechanism to
determine how far along the operation progressed
before failure and, if necessary, how to undo partially
completed operations.
With this information, even across PrivVM failures,
VM management operations
are either executed to completion
or aborted and there is a response to all XenStore requests,
thus leaving the VI in a consistent state.

\subsection{Implementation Complexity}
\label{sec:imp_complexity}

\begin{table}
\tbl{
Implementation complexity of our RVI.
Lines of code (LOC) added or modified to implement
the recovery mechanisms for the three RVI components.
For the VMM and PrivVM,
the ``Enhanced'' mechanism
includes the LOC for all improvements made \textit{in addition}
to the ``Basic'' mechanism.
\label{table:impl_compl}}{
\begin{tabular}{|c|c|c|c|c|}
\hline
Component & Mechanism & User-level & Kernel & VMM \\
\hhline{|=|=|=|=|=|}
\multirow{2}{*}{VMM} & Basic & 0 & 0 & 830 \\
\cline{2-5}
& Enhanced & +0 & +0 & +70 \\
\hline
\multirow{2}{*}{PrivVM} & Basic & 1730 & 1770 & 350 \\
\cline{2-5}
& Enhanced & +575 & +0 & +15 \\
\hline
DVM & Failover & 780 & 1390 & 340 \\
\hline
\end{tabular}
 }
\end{table}

As a partial measure of the engineering effort
required to implement our RVI,
\mbox{Table \ref{table:impl_compl}}
shows the number of lines of code (LOC)
added or modified to implement the recovery
mechanisms for the three RVI components,
broken down by the system layer (privilege level)
at which the new or modified code executes.
The basic PrivVM recovery
mechanism has the highest LOC count.
Most of this code, at the user and kernel levels,
is related to backing up and restoring
the XenStore and XenStored state.
At the VMM level, most of the code is for detecting
PrivVM failures, booting a new PrivVM instance,
and cleaning up the state of the old PrivVM instance.
The added code for the ``enhanced'' PrivVM recovery mechanism
is to reduce inconsistencies following PrivVM recovery,
as described in the last paragraph of Subsection \ref{sec:umprivvm}.

Similarly to the PrivVM, the VMM has state that
must be maintained across a recovery.
However, unlike the PrivVM,
the VMM preserve this state in place, in memory.
This reduces the amount
of code needed for saving and restoring state.
As shown in \mbox{Table \ref{table:impl_compl}},
over 90\% of the additions and modifications are
for implementing the basic VMM microreboot capability
described in Section \ref{sec:basicrec}.
Altogether, the eight enhancements described in
Sections \ref{sec:improve}
and \ref{sec:enhadd} required the addition
or modifications of only \mbox{70 LOC}.

For DVM failover, the modifications
to the VMM implement a mechanism to directly
notify AppVMs that a DVM has failed, thus causing
failover to be initiated quickly.
At the kernel level, more than half of the modifications
are to the network frontend driver in the AppVMs to implement
the failover to the backup DVM.
Most of the remaining changes are to the block device
frontend driver in the AppVMs to forward failure notifications
to the RAID driver \cite{Le13}.
At the user level, the bulk of the changes are for
a dedicated ``DVM manager'' process, that runs in the PrivVM
and is responsible for interacting with
the XenStore and updating device information during failover.

\section{Evaluation of the Resiliency of the RVI}
\label{sec:evalvi}

This section evaluates our complete RVI to determine
how well the various VI detection and recovery mechanisms
work together.
This evaluation uses the 5AppVM configuration
(Figure\ \ref{fig:5appvm}) running on bare hardware,
hosting the LVS workload (Subsection\ \ref{sec:workloads}).
The evaluation involved injecting
between 940 and 1980 register faults during
the execution of each of the three VI components.
Details of the experimental setup are
described in Section \ref{sec:exp}.

Recovery from a VI component failure
is considered \textit{successful} if
(1)
the LVS workload completes successfully ---
no more than two requests from the clients fail
(Subsection \ref{sec:workloads}),
and
(2) at most one AppVM fails and
the recovered VI is able to continue
hosting the remaining VMs as well as create and host new VMs
(Section \ref{sec:vmmfail}).

\begin{figure}
\centerline{\includegraphics[width=2.5in]{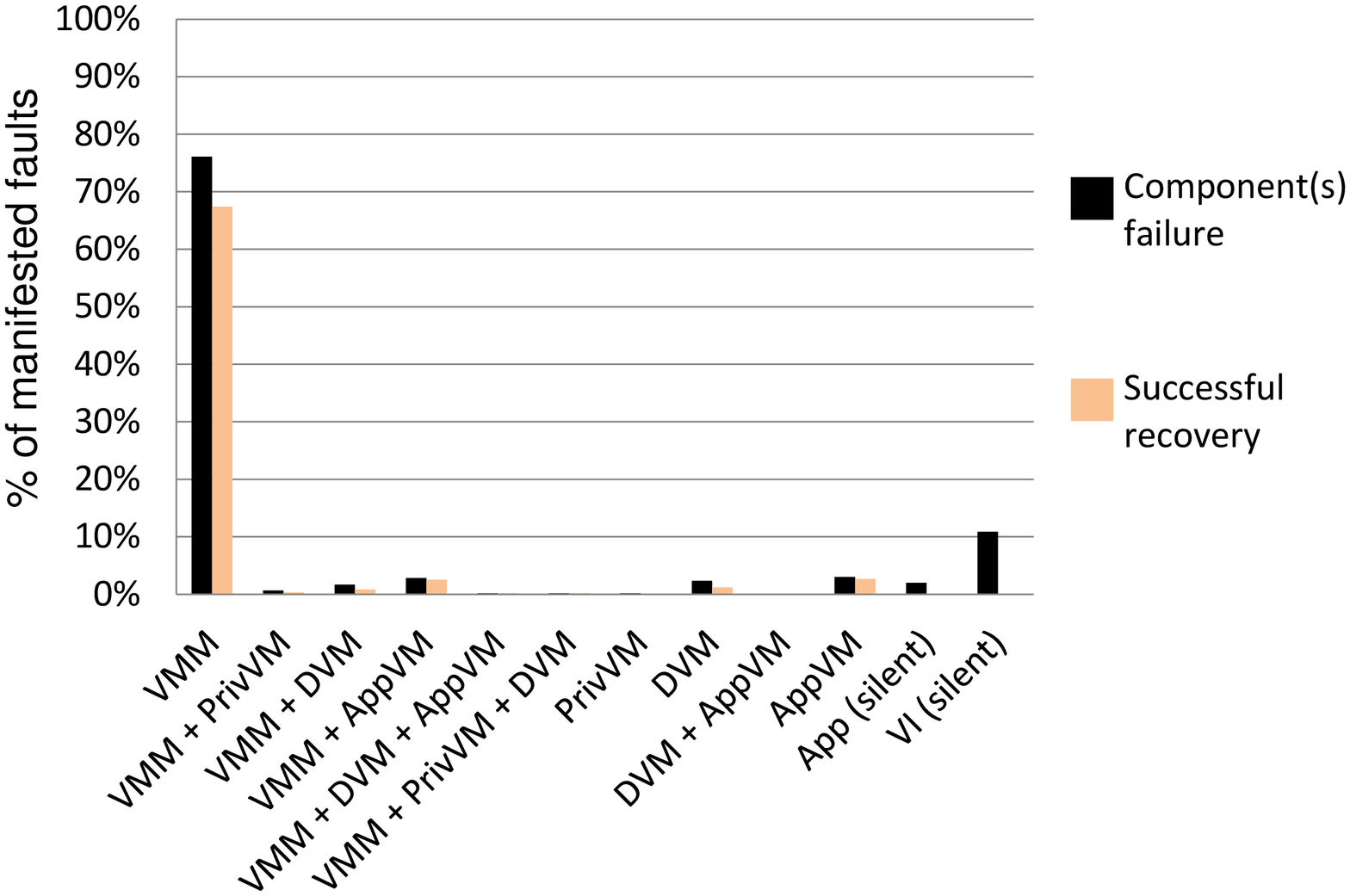}}
\caption{
Distributions of component failures and
successful recovery from those failures
when injecting faults into CPU registers during VMM execution.
}
\label{fig:lvsvmm}

\vspace*{4ex}

\centerline{\includegraphics[width=2.5in]{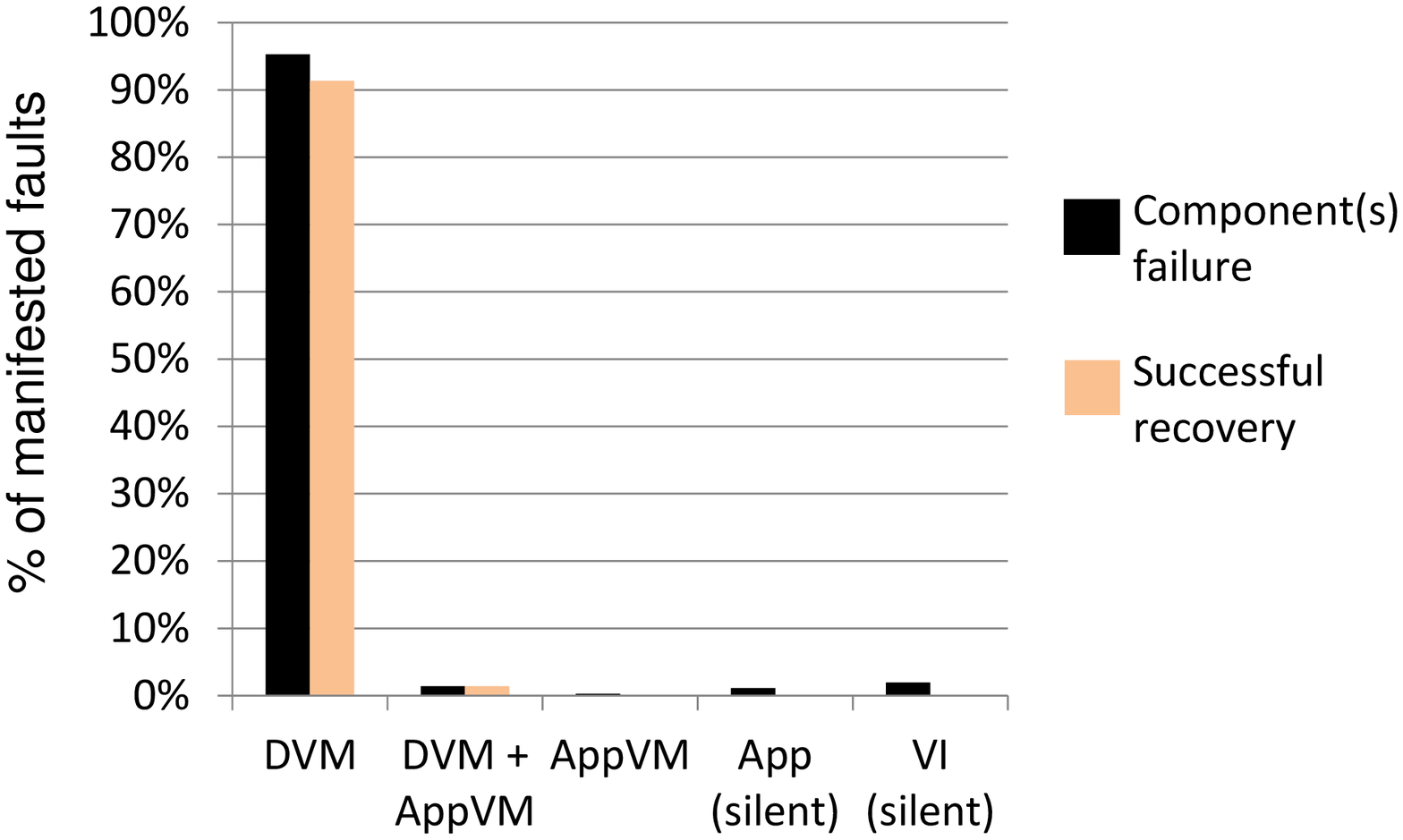}}
\caption{
Distributions of component failures and
successful recovery from those failures
when injecting faults into CPU registers during DVM execution.
}
\label{fig:lvsdvm}

\vspace*{4ex}

\centerline{\includegraphics[width=2.5in]{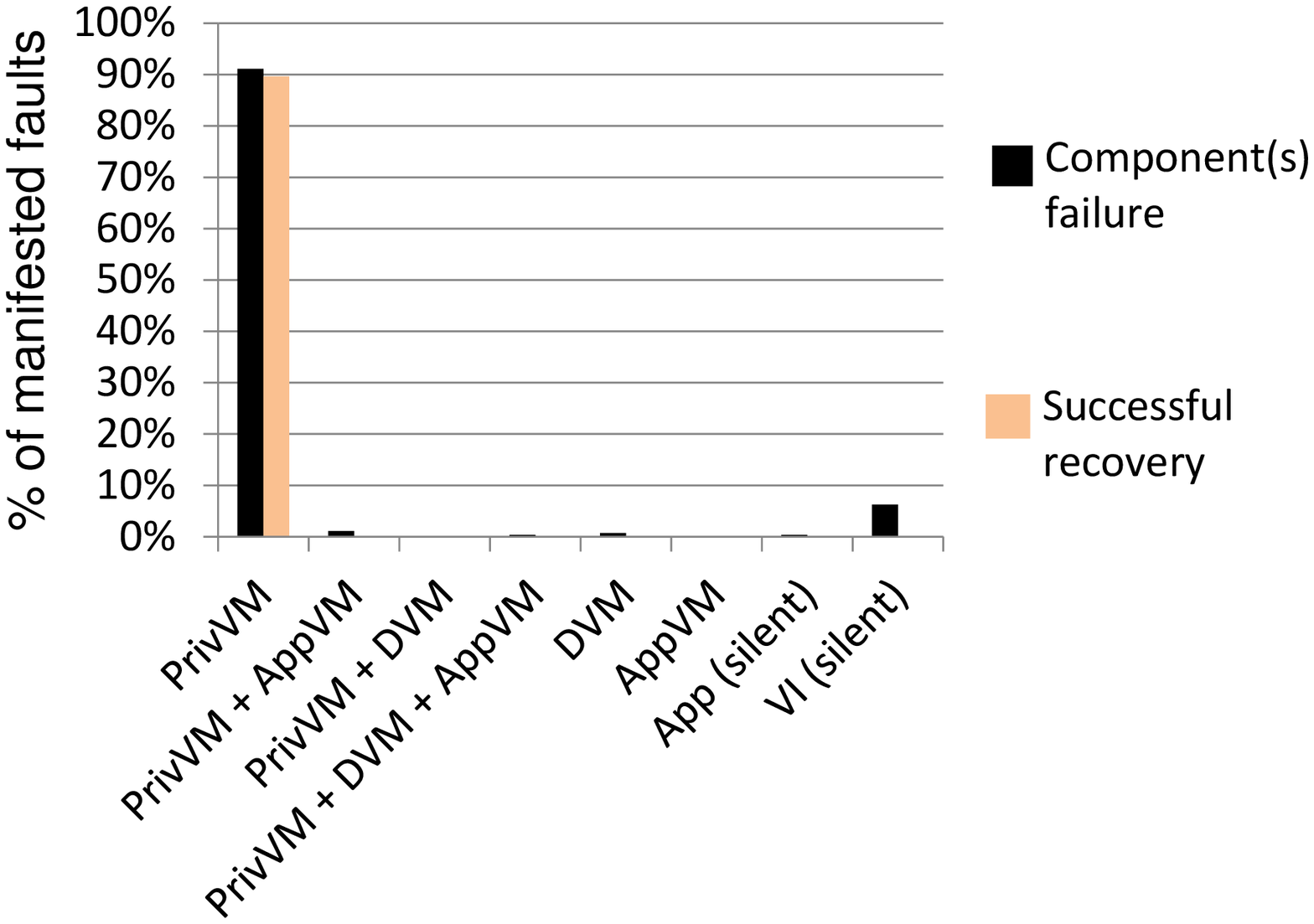}}
\caption{
Distributions of component failures and
successful recovery from those failures
when injecting faults into CPU registers during PrivVM execution.
}
\label{fig:lvsprivvm}
\end{figure}

In most cases, recovery using microreboot is expected to be successful
only if single faults do not manifest as errors
in multiple components of the system.
Hence, for the RVI, it is important to evaluate
the likelihood that a
fault in one VI component can cause
failures in other components of the system.
Figures \ref{fig:lvsvmm}-\ref{fig:lvsprivvm} show
the distributions of component failures caused
by faults injected during the execution
of each one of the VI components,
along with the distributions of successful recoveries
from those failures.
These results show that
the vast majority of
component failures are confined to the component
into which the faults are injected.
This demonstrates that the Xen VI
provides a high degree of fault isolation.
As a result,
as shown in Figures \ref{fig:lvsvmm}-\ref{fig:lvsprivvm},
our RVI recovers successfully from
a great majority of component failures.

While uncommon,
a single fault in a VI component
can cause other components
to fail either together with the faulty component
or independently.
For example, when faults are injected during VMM execution
(Figure \ref{fig:lvsvmm}),
about 2.8\% of manifested faults
cause an AppVM to fail together with the VMM.
Not surprisingly,
given the privileged nature of the VMM,
faults in the VMM cause
the highest rate of multiple component failures
compared to faults in other VI components.
Specifically, with faults occurring during VMM execution,
close to 5.5\% of manifested faults
result in the failure of multiple system components.
With faults occurring during the execution of the DVM or PrivVM,
less than 1.5\% of manifested faults
result in the failure of multiple system components.

Not all failures of multiple components are
due to the direct propagation of errors across components.
Specifically, a fault may initially corrupt
only a single component and cause the failure
of only that component.
However, that failure may be followed by an
incomplete recovery of the affected component,
leaving the system in an
inconsistent state, leading to the failure
of other components.

When there are multiple component failures,
successful system recovery requires
recovery to be performed successively
on each failed component.
For example, for faults occurring during VMM execution,
about 2.5\% of manifested faults
lead to the failure of a DVM and/or PrivVM.
Recovery is successful in only about half of these cases,
requiring VMM recovery followed by DVM and/or PrivVM recovery.
The low recovery success rate in this case
is likely due to incomplete VMM recovery.

As shown in Figures \ref{fig:lvsvmm}-\ref{fig:lvsprivvm},
some faults in VI components
are not detected by our detection mechanisms, but result
in system failures (as defined in Section\ \ref{sec:rvs}).
Specifically, the fault may not be detected but still cause the 
LVS workload to fail (\textit{App-Silent} failures)
or prevent the VI
from correctly hosting current AppVMs
or creating new AppVMs (\textit{VI-silent} failures).
\textit{App-Silent} failures can occur,
for example, when faults in the DVM cause files used
for servicing client requests to become corrupted.
The vast majority of \textit{VI-silent} failures
are caused by faults in the VMM and PrivVM.
The majority those
caused by faults in the VMM
are due to failure of the entire system,
most likely brought on by triple faults
(see Subsection \ref{sec:evalrecfail}).

\begin{table}
\tbl{
For faults in each of the Xen VI components,
the recovery success rates out of
detected failures.
\label{table:lvsres}}{
\begin{tabular}{|c|c|}
\hline
VI Component & Successful Recovery Rate \\
\hhline{|=|=|}
VMM & 87.5\% \\
\hline
DVM & 96.0\% \\
\hline
PrivVM & 96.8\% \\
\hline
\end{tabular}
 }
\end{table}

As a summary of the effectiveness of our recovery mechanisms,
implemented together in one system,
Table \ref{table:lvsres} shows,
for faults in each of the Xen VI components,
the recovery success rate out of detected failures.
Faults in the VMM result in the lowest success rate
then faults in the other VI components.
This is likely due to the fact that ReHype
reuses state from the failed VMM instance.

 \section{Related Work}
\label{sec:related}

\textit{Microreboot} plays a critical role in all
the recovery mechanisms of our RVI.
The term \textit{microreboot} was introduced in \cite{Candea04}
to denote an inexpensive
recovery technique for software systems,
based on recovery of only the failed components.
Many researchers have proposed the use of microreboot
for recovery from device driver failures
\cite{Fraser04,LeVasseur04,Swift05a,Herder07,Le09,Jo10,Le13}.
ReHype is different from these works in that
it microreboots the hypervisor
and addresses the problems of preserving
system components when a lower layer,
the underlying system, is rebooted.

The two works that are most closely related to ReHype
are RootHammer
\cite{Kourai11}
and Otherworld
\cite{Depoutovitch10}.
RootHammer uses microreboot to rejuvenate
\cite{Huang95} virtualized systems based on the Xen VI.
It reduces the time for this rejuvenation
by rebooting only the Xen VMM
and \mbox{Domain 0}, while preserving in memory the states of VMs
and their configurations.
During rejuvenation, \mbox{Domain 0} is properly shut down
and the VMs suspend themselves cleanly.
New instances of the VMM and \mbox{Domain 0} are booted,
without a hardware reset,
and execution of the previously suspended VMs is resumed.

Unlike ReHype, RootHammer performs the VMM (and \mbox{Domain 0})
reboot within a healthy and functioning system,
so that suspension of system components can be done
cleanly, without state corruption.
Hence, RootHammer does not need to resolve potential
inconsistencies within the VMM state, between
the VMM and VMs, and between the VMM and hardware.
Furthermore, with RootHammer, there is no concern
for the safety of the VMM due to corrupted VM states
during VM re-integration.
Unlike RootHammer, ReHype does not require
\mbox{Domain 0} to be rebooted and preserves in place
\mbox{Domain 0} as well as management structures for the AppVMs
across a VMM failure.
As result, the down time with RootHammer is
tens of seconds, much longer than with ReHype
(Section\ \ref{sec:evallatency}).

Otherworld \cite{Depoutovitch10}
allows a Linux kernel to be recovered from failures,
using microreboot, while preserving
in place the state of the running processes.
Otherworld rebuilds many kernel data structures associated
with each process, such as the process descriptor,
the file descriptor table, and signal handler descriptors.
Restoration of kernel components requires
traversing many complex data structures in a possibly
corrupted kernel, increasing the chance of failed recoveries.
In many cases, user-level processes require custom crash procedures
in order to properly resume execution.
In the version of Otherworld evaluated in \cite{Depoutovitch10},
recovery of user-level processes involves copying of
the entire memory state of each process.
As a consequence of the complex recovery procedure,
service interruption time is tens of seconds.
There is mention in \cite{Depoutovitch10} of the possibility
of directly mapping, instead of copying, user-level processes'
memory states, but that option is not evaluated.

Compared to Otherworld,
ReHype benefits from the simplicity of the state that the
VMM keeps for the VMs, enabling
a simpler and faster recovery process.
ReHype reuses the VM descriptors in place and does
not copy VM memory states.
With ReHype, all the states
of the applications are maintained within the VMs.
Hence, application failure handlers or
any other application modifications are not needed.
VM failure handlers could potentially be useful
for performing data integrity checks in the VM using
VM-specific knowledge.
Since there are fewer types of kernels than there are
applications, if VM failure handlers are needed,
fewer have to be written.

Hardware-enforced protection domains can potentially be used
to reduce silent data corruption and increase the
rate of successful recoveries.
In the context of virtualized systems,
this approach is proposed with a mechanism called TinyChecker
in \cite{Tan12}.
TinyChecker relies on nested virtualization to run a small
checker hypervisor beneath the main hypervisor.
Critical VMM data structures used for managing
the system and most of the memory states of VMs
are write protected.
Writes to these protected memory areas
trap to the checker hypervisor.
Based on the context of an access, TinyChecker
determines which memory areas can be safely modified.
For potentially unsafe modifications,
TinyChecker uses on-demand checkpointing
to save a copy of the memory location before
allowing the update.
If a VMM failure occurs,
the checkpoint can be used to restore a valid state.

In \cite{Tan12} there is no evaluation of
TinyChecker's effectiveness for detection or recovery.
TinyChecker incurs overhead for nested virtualization.
Even though TinyChecker itself is likely to be reliable
since it is small and simple, it is susceptible to hardware faults.
TinyChecker, or something like it, is likely to be most
useful \textit{in conjunction} with ReHype.
TinyChecker by itself cannot protect against
erroneous updates of ``valid'' memory areas and cannot
resolve inconsistencies due to recovery.

We initially proposed and evaluated an earlier
version of ReHype in \cite{Le11a}.
To the best of our knowledge, there has been no
other implementation or evaluation, in the context
of recovery from failures, of a mechanism that
can reboot a hypervisor while preserving
hosted AppVMs in place.
We proposed and evaluated our PrivVM recovery
mechanism in \cite{Le12}, and believe that work
to also be unique.
This paper extends our original ReHype work
by incorporating new recovery enhancements,
providing an evaluation and optimization of
recovery latency,
and greatly expanding the recovery effectiveness
evaluation to include
additional fault types, operation on bare hardware,
and hosting of FV VMs.
In \cite{Le12}
we presented an evaluation
of an earlier version of our complete RVI.
This paper presents the first evaluation of our complete
RVI on bare hardware, as it would actually be deployed.
 \section{Conclusions and Future Work}
\label{sec:concl}

Over the last decade,
there has been a rapid increase
in the use of system-level virtualization
in servers and datacenters of all sizes.
The virtualization infrastructure (VI) is a software
layer that allows multiple VMs to run on a single host.
The VI introduces a critical single point of failure where
a transient hardware fault or software fault that cause
a VI component to fail, lead to the failure of all
the VMs running on the host.
This paper presented a resilient VI (RVI), based on Xen,
that uses microrollback to recover failed VI
components, while preserving VMs running applications (AppVMs)
in place.
Due to the critical role in the VI of the hypervisor (VMM),
much of the paper is focused on ReHype -- the recovery
mechanism for the hypervisor.

ReHype was developed by first implementing
the basic recovery functionality and then
incrementally enhancing the basic mechanism,
guided by fault injection results.
For successful recovery, ReHype must avoid or
resolve state corruption and inconsistencies in
the hypervisor, between hypervisor and VMs, and
between the hypervisor and hardware.
A fault injection campaign emulating
a variety of software and transient hardware faults,
yielded a successful recovery rate of approximately 88\%.
This is achieved with essentially no performance overhead
during normal operation, negligible memory overhead,
and changes or additions of only 900 lines of code in
the hypervisor.
With ReHype, recovery latency --
the duration service interruption for the AppVMs --
is less than 1\ sec.

We have also developed recovery mechanisms for the
driver VMs (DVMs) and privileged VM (PrivVM).
These mechanisms, together with ReHype, form our RVI.
We evaluated the RVI hosting a workload
consisting of cluster middleware and web servers.
For this entire system,
for detected faults in the VMM, PrivVM, and DVM,
the failure recovery rate were 87\%, 96\%, and 96\%, respectively.

ReHype uses simple detection mechanism that detect
crashes and hangs.
As a result, depending on the fault type
and system setup, between 10\% and 36\% of manifested faults
lead to \textit{silent failures} --
they are not detected and thus the recovery mechanisms are
not triggered.
Two key goals for future research are:
reducing silent failures,
and further improvements of the recovery success rate.
For both goals, hardware-enforced protection domains
may be used to reduce random memory corruption.
Additional benefits could be gained by
maintaining redundant information during normal operation
and proactively performing sanity checks.
Such redundant information could also be used
to transactionalize
certain hypercalls and provide reliable
re-delivery of pending virtual interrupts.
 \bibliographystyle{plain}
\bibliography{rehypej_bib}

\end{document}